\documentclass[twocolumn,english,prl,superscriptaddress]{revtex4}
\usepackage[T1]{fontenc}
\usepackage[latin9]{inputenc}
\usepackage{float}
\usepackage{amsmath}
\usepackage{graphicx}
\usepackage{parskip}
\usepackage{babel}
\usepackage{comment}
\usepackage{notes2bib}
\usepackage{indentfirst}

\makeatletter
\@ifundefined{textcolor}{}
{%
	\definecolor{BLACK}{gray}{0}
	\definecolor{WHITE}{gray}{1}
	\definecolor{RED}{rgb}{1,0,0}
	\definecolor{GREEN}{rgb}{0,1,0}
	\definecolor{BLUE}{rgb}{0,0,1}
	\definecolor{CYAN}{cmyk}{1,0,0,0}
	\definecolor{MAGENTA}{cmyk}{0,1,0,0}
	\definecolor{YELLOW}{cmyk}{0,0,1,0}
}

\makeatother

\begin{document}
		
\title{Topology detection in cavity QED}
\author{Beatriz Pérez-González}
\affiliation{Instituto de Ciencia de Materiales de Madrid (ICMM, CSIC)}
\author{Álvaro Gómez-León}
\affiliation{Instituto de Física Fundamental (IFF, CSIC)}
\author{Gloria Platero}
\affiliation{Instituto de Ciencia de Materiales de Madrid (ICMM, CSIC)}

	\begin{abstract}
		We explore the physics of topological lattice models in c-QED architectures for arbitrary coupling strength, and the use of the cavity transmission as a topological marker. For this, we develop an approach combining the input-output formalism with an expansion in quantum fluctuations which allows to go beyond the small-coupling regime. We apply our formalism to a fermionic Su-Schrieffer-Heeger (SSH) chain coupled to a single-mode cavity, and find that the cavity can indeed act as a quantum sensor for topological phases, where the initial state preparation plays a crutial role. Additionally, we discuss the persistence of topological features as the coupling strength increases, in terms of an effective Hamiltonian, and calculate the entanglement entropy. 
	\end{abstract}

\maketitle
	
\textit{Introduction.} $-$ Cavity Quantum Electro-Dynamics (c-QED) studies the interaction between
light and matter at the most elementary level, either with individual
atoms \cite{AtomicQED} or solid-state devices \cite{CavityQEDAtomToCond}. These hybrid systems have revealed themselves as an important tool for the control of quantum systems \cite{ManipulatingIntertwined}, and an essential landmark in the development
of quantum technologies \cite{QEDarchQC,DecQCCavQED, QInfoCircuitQED}.
This is because the coherent interaction of both systems allows for
an efficient transfer of information between the two \cite{StateTransfer, QSTinDistantNodes},
provided that their coupling is larger than the
losses in the system. 

A step further can be taken by considering the combination of quantum
light and complex quantum materials,
such as topological systems \cite{TopCav,TopCav2,MajoranaDet,cQEDwithTopSup}.
A celebrated example of such are topological insulators (TIs), which have mid-gap states which are exponentially localized at the boundaries and protected by certain symmetries of the band structure. 

In this work, we study the interplay of fermionic topological systems and c-QED architectures. Our aim is to use the cavity transmission as a topological marker, identifying the experimental signatures of non-trivial topological properties. For that purpose, we develop a framework combining input-output theory with an expansion in quantum fluctuations whose validity extends to arbitrary coupling strength. Finally, we study in detail the case of the SSH chain \cite{mipaper1,mipaper2,doublonMiguel} coupled to a single mode cavity. 

\textit{Mean-field and Fluctuations Hamiltonian.} $-$
We consider the Hamiltonian $H = \Omega d^\dagger d + H_S + V$. The first term describes a cavity with frequency $\Omega$, being $d$ ($d^\dagger)$ the destruction (creation) photon operator, while $H_S$ corresponds to the single-particle Hamiltonian of a topological system. The interaction term has the form $V = g(d^\dagger + d)Z $, being $Z$ the fermionic coupling operator and $g$ the coupling strength.

We apply a mean-field (MF) decoupling and write each operator in
$V$ as its average plus the fluctuations around it: $\mathcal{O}=\langle\mathcal{O}\rangle+\mathcal{\delta O}$,
being $\mathcal{\delta O}$ the fluctuations operator. Then, we identify two contributions: i) the MF Hamiltonian, which is linear in fluctuations and can be described by $\tilde{H}_S = H_S + g\left(\langle d^{\dagger}\rangle+\langle d\rangle\right)Z$ for fermions, and $H_\Omega = \Omega d^{\dagger}d+g\langle Z\rangle\left(d^{\dagger}+d\right)$ for photons; and ii) the quadratic part in fluctuations $H_\delta = g\left(\delta d^{\dagger}+\delta d\right)\delta Z$. Note that the MF part is written in terms of the original operators using $\delta \mathcal{O} = \mathcal{O} - \langle O \rangle $, and that constant terms have been neglected.




Finally, a complete characterization requires self-consistent solutions to $\langle Z\rangle$,
$\langle d^{\dagger}\rangle$ and $\langle d\rangle$, which can be obtained, to lowest order, from the MF Hamiltonian. This requires to rotate $H_\Omega$ with $R = \exp\left\{ -g\langle Z\rangle\left(d^{\dagger}-d\right)/\Omega\right\} $,
to find:

\begin{equation}
\tilde{H}_{\Omega}=R^{\dagger}H_{\Omega}R=\Omega d^{\dagger}d-\frac{g^{2}\langle Z\rangle^{2}}{\Omega}. \label{eq:Photon-rotated}
\end{equation}

Importantly, $\tilde{H}_\Omega$ is now diagonal in the rotated basis, where it is easy to determine the average of the photon operators: $\langle d^{\dagger}\rangle+\langle d\rangle=-2g \langle Z\rangle/\Omega$. We can then write $\tilde{H}_S$ in terms of fermionic averages only:

\begin{equation}
\tilde{H}_{S}=H_{S}-2\frac{g^{2}}{\Omega}\langle Z\rangle Z.\label{fermionicMF}
\end{equation}

Notice that the interaction with the cavity photons introduces a term proportional to $Z$, that can affect the topological properties of $\tilde{H}_S$. Fortunatelly, Eq. \ref{fermionicMF} simplifies the calculation of $\langle Z\rangle$,
which can now be obtained using an iterative numerical procedure involving fermionic degrees of freedom only. Finally, $\tilde{H}_\delta$ can also be written in the rotated frame:

\begin{equation}
\tilde{H}_{\delta}=R^{\dagger}H_{\delta}R=g\left(d^{\dagger}+d\right)\sum_{\vec{\alpha}}\tilde{Z}_{\vec{\alpha}}\tilde{X}^{\vec{\alpha}},
\end{equation}

where $\tilde{Z}_{\vec{\alpha}}=\langle\tilde{\alpha}_{1}|\left(Z-\langle Z\rangle\right)|\tilde{\alpha}_{2}\rangle$, and we define the Hubbard operators $\tilde{X}^{\vec{\alpha}}=\tilde{X}^{\alpha_{1},\alpha_{2}}=|\tilde{\alpha}_{1}\rangle\langle\tilde{\alpha}_{2}|$,
with $|\tilde{\alpha}_{i}\rangle$ ($\alpha_{i}=1,\ldots,N$) being
the eigenstates of  $\tilde{H}_{S}|\tilde{\alpha}_{i}\rangle=\tilde{E}_{\alpha_{i}}|\tilde{\alpha}_{i}\rangle$,
and $\vec{\alpha}=\left(\alpha_{1},\alpha_{2}\right)$. \\
Indeed, $\tilde{Z}_{\vec{\alpha}}$ measures the effect of fluctuations on the MF eigenstates: it vanishes in the two asymptotic limits $g\ll\Omega,\tilde{E}_{\vec{\alpha}}$
and $g\gg\Omega,\tilde{E}_{\vec{\alpha}}$, implying that fluctuations are completely suppressed in the rotated frame, where the two systems effectively decouple, making the MF description exact. Away from these limits, fluctuations are relevant and must be estimated. \\

\textit{Calculation of the cavity transmission.} $-$ To study the cavity transmission $t_c$, we include the coupling of the cavity photons to the external modes \cite{QuantumNoise}. By means of input-output theory, we can obtain the input $\tilde{b}_{\mathrm{in},l}$ and output $\tilde{b}_{\mathrm{out},l}$ fields at each of the sides $l=1,2$ of the cavity, and write $t_c = |t_c| e^{i\varphi} = \langle \tilde{b}_{\mathrm{out}}\rangle /  \langle \tilde{b}_{\mathrm{in}}\rangle$ \cite{Note1}.

The starting point is the Langevin equation of motion (EoM) for the cavity field $d(t)$ \cite{QuantumNoise,QEDarchQC}:

\begin{equation}
\partial_{t}d(t)  =  -i\left(\Omega-i\frac{\kappa}{2}\right)d(t) - \sum_{l=1,2}\sqrt{\kappa_{l}}\tilde{b}_{\mathrm{in},l}(t)  - ig\sum_{\vec{\alpha}}\tilde{Z}_{\vec{\alpha}}\tilde{X}^{\vec{\alpha}}(t) ,
\label{eq:EOMd}
\end{equation}

where $\kappa_l$ represents the coupling between the cavity and the outside modes, and $\kappa=\kappa_{1}+\kappa_{2}$. To solve it, we consider the retarded Green function $G(t,t^\prime) = -i\theta(t-t^\prime)\langle [d(t),d^\dagger (t^\prime)]\rangle $. Considering its Langevin EoM with local dissipation $\kappa$, obtained from the quantum regression theorem, we notice that the EoM for $G(t,t^\prime)$ is analogous to Eq. \ref{eq:EOMd}, but without the term $\sqrt{\kappa_l} \tilde{b}_\mathrm{in}$. Then, it is only required to notice that $G(t,t^\prime)$ is the resolvent of Eq. (\ref{eq:EOMd}), which means we can write a general solution for $d(\omega)$ as $d(\omega) = -iG(\omega)\sum_{l=1,2}\sqrt{\kappa_l}\tilde{b}_{\mathrm{in},l}(\omega)$ for arbitrary $g$ \cite{Note1}. Together with the input-output relation $\tilde{b}_{\mathrm{out},l}(t) = \tilde{b}_\mathrm{in,l}(t) + \sqrt{\kappa_l}d(t)$ (obtained from the reversed-time EoM for $d(\omega)$), this result let us write 

\begin{equation}
t_c(\omega) = \frac{\langle\tilde{b}_{\mathrm{out},2}\rangle}{\langle\tilde{b}_{\mathrm{in},1}\rangle} = -i\sqrt{\kappa_1 \kappa_2} G(\omega),
\label{eq:Exact-transmission}
\end{equation}

To connect with the standard input-output result for $t_c$, the EoM for $G(\omega)$ can be solved, using

\begin{equation}
\partial_{t}\tilde{X}^{\vec{\alpha}}(t) = i\left(\tilde{E}_{\vec{\alpha}} - i\frac{\gamma}{2}\right)\tilde{X}^{\vec{\alpha}}(t) + ig\left[d^{\dagger}(t) + d(t) \right]Y_{\vec{\alpha}}^{-}(t)
\end{equation}

where $Y_{\vec{\alpha}}^{-}(t) = \sum_\beta (\tilde{Z}_{\alpha_2 \beta}\tilde{X}^{\alpha_1 \beta}(t) - \tilde{Z}_{\beta \alpha_1}\tilde{X}^{\beta \alpha_2}(t))$, $\tilde{E}_{\vec{\alpha}} = \tilde{E}_{\alpha_{1}}-\tilde{E}_{\alpha_{2}}$, and $\gamma$ is the spectral broadening. To close the system of equations, a decoupling scheme is required. For this, we notice that contributions from fluctuations are small for $g\ll\Omega,\tilde{E}_{\vec{\alpha}}$
and $g\gg\Omega,\tilde{E}_{\vec{\alpha}}$, when working in the rotating frame. Under these conditions, we can rewrite $d^{\left(\dagger\right)}\left(t\right)\tilde{X}^{\vec{\alpha}}\left(t\right)\approx\langle \tilde{X}^{\vec{\alpha}} \rangle d^{\left(\dagger\right)}\left(t\right) + \langle d^{\left(\dagger\right)}\rangle \tilde{X}^{\vec{\alpha}}(t)$. Note that we are also neglecting extra correlation terms between photonic and fermionic operators by assuming that fluctuations are small, which would only be relevant near resonances. For the same reason, $\langle d^{(\dagger)}\rangle $ and $\langle \tilde{X} \rangle $ can be calculated using $\tilde{H}_\Omega$ and $\tilde{H}_S$, respectively. As a result, $\langle d^{(\dagger)} \rangle \sim 0$ \textit{in the rotated frame}. We will also neglect $\langle \tilde{X} \rangle d^\dagger (t)$ for being small in the regime of interest \cite{PRASigmund}.

With these approximations, we can solve the EoM for $G(\omega)$ and obtain an analytical expression for $t_c$:

\begin{equation}
	t_{c}=\frac{\langle\tilde{b}_{\mathrm{out},2}\rangle}{\langle\tilde{b}_{\mathrm{in},1}\rangle}=\frac{i\sqrt{\kappa_{1}\kappa_{2}}}{\Omega-\omega+g^{2}\tilde{\chi}\left(\omega\right)-i\frac{\kappa}{2}}\label{eq:Transmission}
\end{equation}

where $\tilde{\chi}\left(\omega\right)$ is the electronic susceptibility, 

\begin{equation}
	\tilde{\chi}\left(\omega\right)=\sum_{\vec{\alpha}\beta}\tilde{Z}_{\vec{\alpha}}\frac{\tilde{Z}_{\alpha_{2},\beta}\langle\tilde{X}^{\alpha_{1},\beta}\rangle-\tilde{Z}_{\beta,\alpha_{1}}\langle\tilde{X}^{\beta,\alpha_{2}}\rangle}{\omega + \tilde{E}_{\vec{\alpha}} - i\frac{\gamma}{2}} \label{eq:susceptibilityext}
\end{equation} 

with $\langle\tilde{X}^{\vec{\alpha}}\rangle=\delta_{\alpha_{1},\alpha_{2}}p_{\alpha_{1}}$,
being $p_{\alpha}$ the occupation of state $|\tilde{\alpha}\rangle$ in the
fermionic density matrix $\rho_{f}=\sum_{\alpha}p_{\alpha}\tilde{X}^{\alpha,\alpha}$. Eq. \ref{eq:susceptibilityext} is analogous to the electronic susceptibility for the small-$g$ regime in the standard input-output formalism \cite{PRASigmund}, but in this case all fermionic parameters are renormalized by the MF self-consistency equations.  This allows to extend the validity of Eq. \ref{eq:Transmission} to the very-large-$g$ regime, but we will see below that it also reproduces the behaviour in the intermediate regime accurately.


\textit{Results.} $-$ The theoretical framework developed so far can be used for an arbitrary $H_\mathrm{S}$ and $Z$. Now we apply our formalism to a particular fermionic 1D topological system described by $H_{S}=\sum_{\langle i,j\rangle=1}^{N}t_{ij}c_{i}^{\dagger}c_{j}$, where $t_{ij}$ is the hopping amplitude and $c_{i}$ ($c_{i}^{\dagger}$) is the destruction (creation) operator for a spinless fermion at site
$i$ $(i=1,...,N)$. Particularizing for the SSH model \cite{mipaper1,mipaper2,doublonMiguel}, we define the intra- and inter-dimer
hopping amplitudes $t=t_{0}\left(1+\delta\right)$ and $t^{\prime}=t_{0}\left(1-\delta\right)$,
respectively, with $\delta\in\left[-1,1\right]$. This bipartite structure gives rise to two distinct topological phases: the trivial
phase ($\delta<0$), and the topological phase ($\delta>0$), hosting a pair of topologically protected
edge states. In the following, we set $t_0 \equiv 1$ as the energy scale for $\Omega$ and $g$.

For the interaction between the cavity and the chain we consider the dipolar coupling, $Z=\sum_{i}^{N}x_{i}c_{i}^{\dagger}c_{i}$ \cite{ElectronPhotonMesoscopic}, being $x_{i}$ the position in the lattice such that $x_{N}=-x_{1}$. 
We have calculated $\langle Z \rangle $ vs $g$ self consistently and found that for small $g$, $\langle Z \rangle = 0$ (see details in \cite{Note1}). This confirms that the eigenstates of $\tilde{H}_{S}$ coincide with those of
$H_{S}$ ($\tilde{E}_{\vec{\alpha}}\rightarrow E_{\vec{\alpha}}$,
$\tilde{X}^{\vec{\alpha}}\rightarrow X^{\vec{\alpha}}$) and $\tilde{H}_{\delta}$
turns into the original $V$ (i.e. $\tilde{Z}\rightarrow Z$). 

In this regime, $Z_{\vec{\alpha}}$ mediates transitions between all eigenstates of the chain, with the exception of the edge states in the topological phase, which exponentially suppress their coupling with the bulk states as a function of the chain length. In consequence, if the system is initially prepared in its ground state, the interaction with the electronic system will shift the cavity frequency equally for both phases, which means that $t_c$ cannot be used as a topological marker in this regime, and only the initial preparation in an edge state would allow to detect topology \cite{Note1}. In contrast, $\langle Z \rangle \neq 0$ for $g\gg\Omega,\left|t_{i,j}\right|$, accounting for the polarization of the system. However, the global shift in the energy of the cavity photons does not affect the measurement of $t_c$, so we expect $|t_c|\simeq 1$ again.  


\begin{figure}[t!]
	\centering
	\includegraphics{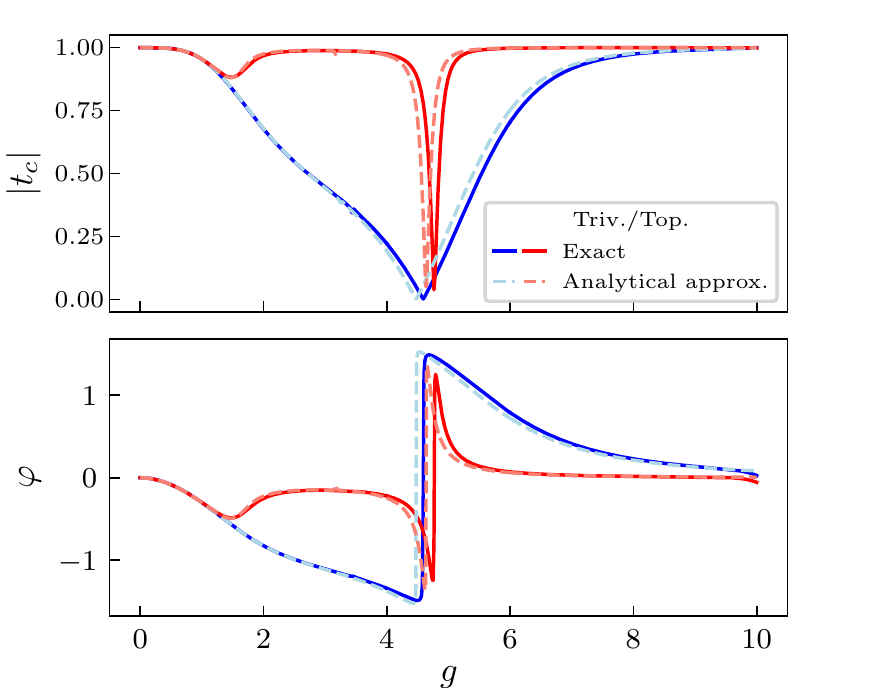}
	\caption{\label{fig:fluctvsexact} $\left|t_{c}\left(\Omega\right)\right|$ and $\varphi$ vs $g$. Dashed lines correspond to the analytical approximation (Eq. \ref{eq:Transmission}), and solid lines to the exact solution (Eq. \ref{eq:Exact-transmission}). The ground state of the chain is occupied. Parameters: $\omega = \Omega = 10, \delta = \pm 0.6$, $N=20$, $\gamma = \kappa_1 = \kappa_2 = 0.01$.}
\end{figure}

All these features are shown in Fig. \ref{fig:fluctvsexact}, where we have calculated $t_c$ at $\omega = \Omega=10$ (highly detuned from the electronic system) for both phases, as a function of $g$, including from the small to the very-large $g$ regime. We consider \textit{the chain with one particle in its lowest energy state}. For small $g$, the peak of maximum transmission, initially found at $\omega = \Omega$, shifts due to the interaction for both phases. The edge states are transparent to the bulk states and their presence is not revealed in $t_c$. However, Eqs. \ref{eq:Transmission} and \ref{eq:Exact-transmission} predict that the break-down of the small-$g$ regime brings essential differences between phases: while the trivial phase decays monotonically until reaching a minimum, the topological phase remains mostly unaffected, except for a notorius dip in $|t_c|$ (which corresponds to a change of sign in $\varphi$). 
Thus, the effect of the topological edge states is not washed away by the chiral-symmetry breaking, and the difference between phases can still be detected. Finally, both phases display perfect transmission when $g$ is sufficiently large, in accordance with the MF analysis. Importantly, the comparison between Eqs. \ref{eq:Transmission} and \ref{eq:Exact-transmission} gives a perfect agreement between the exact and the analytical curves for arbitrary $g$, which means that Eq. \ref{eq:Transmission} captures the behaviour of the system for arbitrary coupling.

\textit{Effective Hamiltonian.} $-$ To explore in more detail the different interaction regimes, we derive an effective Hamiltonian using a Schrieffer-Wolff transformation in the basis of eigenstates of the MF Hamiltonian, where $\tilde{H}_{\delta}$ is considered the perturbation to $\tilde{H}_{S}+\tilde{H}_{\Omega}$. The transformation is defined through the generator $S$ such that $\bar{H}=e^{S}\tilde{H}e^{-S}\simeq\tilde{H}_{S}+\tilde{H}_{\Omega}+\frac{1}{2}\left[S,\tilde{H}_{\delta}\right]$, yielding \cite{Note1}

\begin{eqnarray}
\bar{H} & \simeq & \sum_{i,j}\tilde{E}_i \tilde{X}^{ii}+ \bigg[ \Omega + g^2 \sum_{\vec{\alpha}} \tilde{\Omega}_{\vec{\alpha}} \tilde{Y}^{-}_{\vec{\alpha}} \bigg] d^\dagger d  \nonumber\\
& & + \frac{g^2}{2}\sum_{\vec{\alpha}} \tilde{Z}_{\vec{\alpha}} \sum_\beta \left[ \frac{ \tilde{Z}_{\alpha_2 \beta} \tilde{X}^{\alpha_1 \beta}}{\tilde{E}_{\vec{\alpha}} - \Omega} - \frac{\tilde{Z}_{\beta \alpha_1} \tilde{X}^{\beta \alpha_2} }{\tilde{E}_{\vec{\alpha}} + \Omega} \right] 
 \label{eq:S-W1}
\end{eqnarray}

with $\bar{\Omega}_{\vec{\alpha}} = \tilde{Z}_{\vec{\alpha}}\tilde{E}_{\vec{\alpha}}/(\tilde{E}_{\vec{\alpha}}^2 - \Omega^2 )$. To derive Eq.\ref{eq:S-W1} we have neglected the small correction
provided by two-photon transitions in the rotated frame. Then, $\bar{H}$ includes a shift in the cavity frequency $\Omega$, proportional to $\tilde{\Omega}_{\vec{\alpha}}$, that depends on the state of the electronic system through $\tilde{Z}_{jn}\langle \tilde{X}^{in} \rangle - \tilde{Z}_{ni} \langle \tilde{X}^{nj}\rangle $, as well as a correction to the MF electronic energies (second line). 

\begin{figure}[t!]
	\centering
	\includegraphics{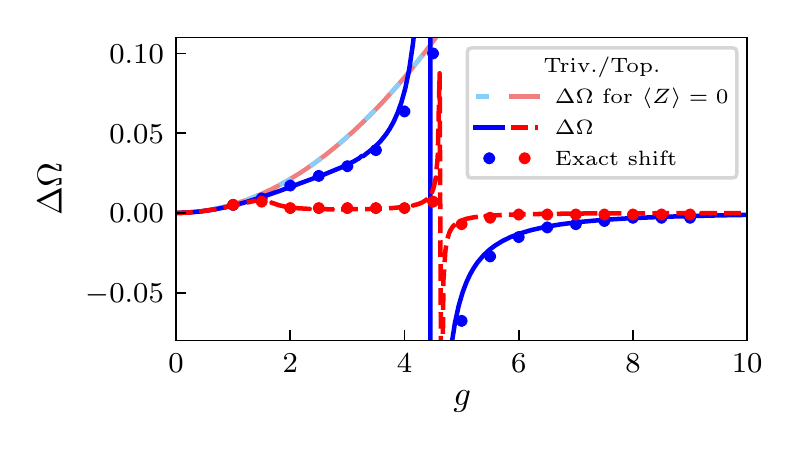}
	\caption{\label{fig:Plotcavityshift} 
		Plot of $\Delta \Omega$ (Eq. \ref{eq:cavityshift}) vs $g$ for both the trivial (blue) and topological (red) phase. The comparison with $\Delta \Omega $ calculated with $\langle Z \rangle$ has been included in lighter colors. The dots corresponds to the exact shift (Eq.\ref{eq:Exact-transmission}). Parameters: $\Omega = 10, \delta = \pm 0.6$, $N=20$}
\end{figure}

The total frequency shift can be obtained from $\Delta \Omega = g^2 \sum_{ijn} \tilde{\Omega}_{ij} \left( \tilde{Z}_{jn}\langle \tilde{X}^{in} \rangle  - \tilde{Z}_{ni} \langle \tilde{X}^{nj}\rangle \right)$, where the expected values are calculated to lowest order using the ground state of $\tilde{H}_S$. In this case, $\Delta \Omega$ reduces to:

\begin{equation}
\Delta \Omega = g^2 \sum_j |\tilde{Z}_{0j}|^2\frac{ \tilde{E}_0 - \tilde{E}_j}{(\tilde{E}_0 - \tilde{E}_j)^2 - \Omega^2}.
\label{eq:cavityshift}
\end{equation}

Fig. \ref{fig:Plotcavityshift} shows $\Delta \Omega$ vs $g$ for the same parameters as Fig. \ref{fig:fluctvsexact}. Dots correspond to the exact shift obtained from Eq.\ref{eq:Exact-transmission}, exhibiting an excellent agreement with the predictions from $\bar{H}$ for arbitrary $g$. For larger $g$, Eq. \ref{eq:cavityshift} also captures the presence of a divergence (signaling a resonance between photons and fermions), in accordance with the drop in $t_c$ shown in Fig. \ref{fig:fluctvsexact} for both phases. Additionally, $\Delta \Omega$ obtained for $\langle Z \rangle = 0$ ($\tilde{Z}\rightarrow Z$, $\tilde{E}_i\rightarrow E_i$) has been included. These results indicate that: i) the dependence of $\tilde{H}_S$ on $\langle Z \rangle$ makes Eqs. \ref{eq:Exact-transmission} and \ref{eq:S-W1} non-perturbative, since $\langle Z \rangle$ is also a function of the other parameters; and ii) the polarization of the system with $\langle Z \rangle \neq 0 $ triggers the onset of the qualitatively distinct behaviour between both phases (Fig. \ref{fig:fluctvsexact}) and the breakdown of the small-$g$ regime, enabling topological detection for equilibrium configurations (i.e. ground state occupied). Also note in Fig. \ref{fig:Plotcavityshift} that $\Delta \Omega$ changes sign after the divergence. Finally, for $g \gtrsim \Omega$, $\delta \Omega \rightarrow 0$ for both phases, as expected: fluctuations are suppressed and there is a global shift in the energy of the photons (Eq. \ref{eq:Photon-rotated}), not detected in $t_c$. 

The change in the eigenenergies of the system due to the interaction is not trivial. The polarization of the system with $\langle Z\rangle \neq 0$ introduces an on-site energy (Eq. \ref{fermionicMF}), which localizes the states and breaks the original chiral symmetry of the SSH model. However, the breaking of the symmetries that provide for topological protection is expected even for small $g$ due to the term in the second line of Eq. \ref{eq:S-W1}. Indeed, as $g$ increases, the topological edge states reduce their energy and eventually penetrate into the bulk band. Despite this, their presence can be detected in $t_c$, even after their disappearance, thus accounting for the differences in $t_c$ between phases. 
Besides, this correction to the electronic MF eigenstates can be used to measure the effect of fluctuations on the ground state of $\tilde{H}_S$. A numerical estimation reveals that it is very small even in the intermediate regime, which explains why there is such a good agreement between the exact (Eq. \ref{eq:Exact-transmission}) and analytical (Eq. \ref{eq:Transmission}) $t_c$ for arbitrary $g$ Fig. \ref{fig:fluctvsexact}.

\textit{Entanglement entropy.} $-$ To gain further insight into the topological features of
the intermediate coupling regime, we explore the entanglement entropy $S_{A}=-\mathrm{tr}_{A}\rho_{A}\ln\rho_{A}$,
where $\rho_{A}$ corresponds to the reduced density matrix of a subsystem
$A$. When the system is divided in two partitions $A\oplus B$, $S_{A}$
measures the amount of quantum correlation between them (note that
$S_{A}=S_{B},$ with $S_{B}$ defined analogously). 

\begin{figure}[t!]
	\centering
	\includegraphics{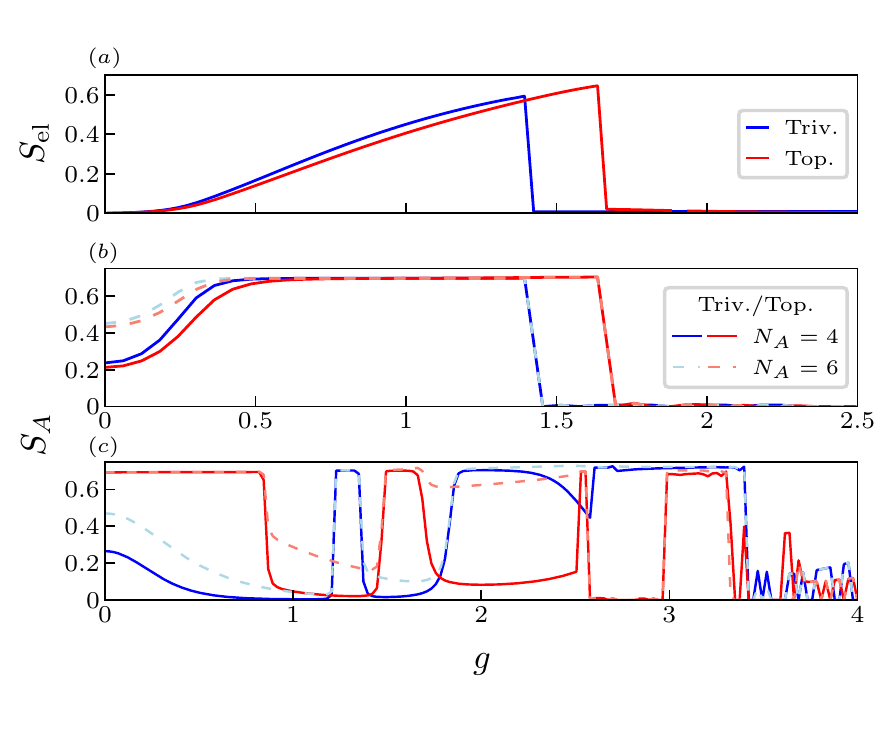}
	\caption{\label{fig:entanglement-entropy}
		\textbf{(a)} $S_{\mathrm{el}}$ vs $g$, for the ground state 
		\textbf{(b)} $S_{A}$ vs $g$ for different partitions $N_{A}$ and the system in its ground state. 
		\textbf{(c)} $S_{A}$ for the system in its $N/2$-th state. 
		Parameters: $\Omega=10$, $\left|\delta\right|=0.6$, and $N=20$}
\end{figure}

We can partition the system in two different ways. First,
we consider a partition separating the fermionic chain and the cavity,
and calculate $S_{\mathrm{el}}$ by tracing out the photonic degrees
of freedom (Fig. \ref{fig:entanglement-entropy}(a)). $S_{\mathrm{el}}$ grows with $g$ and suddendly
drops at a critical coupling value, which is different for
the trivial and topological phase and reproduces the phase transition
captured by the order parameter $\langle Z\rangle$ \cite{Note1}. The alternative partition requires to first integrate out the photonic degrees of freedom. The resulting density matrix for the fermionic
chain encodes the role of photons and can be divided in two parts
$A=\{1,2,...,N_{A}\}$ and $B=\{N_{A}+1,N_{A}+2,...,N\}$ (with $N_{A}\neq N/2$)
of which the entropy can be calculated. From a
topological perspective, the entanglement between these partitions
has been already studied in non-interacting systems and can differentiate
between topological phases \cite{entanglementBerry}.

In Fig. \ref{fig:entanglement-entropy}(b) we plot $S_A$ for the system in its ground state. Its behaviour is very similar to that of $S_\mathrm{el}$ and can be used to obtain information about the phase diagram for
$\langle Z \rangle $. The saturation to $\log2$ indicates
that the ground state becomes a cat state for finite $g$, which is
destroyed when the system polarizes (i.e., $\langle Z\rangle\neq0$),
turning into a fully localized state.


In contrast, Fig. \ref{fig:entanglement-entropy}(c) considers the
$N/2$-th state occupied. This state coincides
with the edge state in the topological phase and with the top of the
valence band in the trivial phase for the isolated chain. The topological phase displays a quantized value $S_{A}=\log2$
for small $g$, independent of the partition, indicating the presence of a maximally entangled state between the boundaries of the system \cite{entanglementBerry}. We can see that the topological
boundary mode is present until the entropy drops and the entanglement for the $N/2$-th state is destroyed, coinciding with the first anti-crossing of the edge state with a bulk state \cite{Note1}.

The persistence of the $\mathrm{log}\left(2\right)$ value for $S_{A}$
in the topological chain for small $g$ means the interaction preserves
the entanglement between ending sites created by the original topological
boundary modes, though their energy and localization length does change
due to the symmetry breaking. This can have
important implications for designing quantum information protocols
with c-QED structures in which correlation between distant sites needs to be exploited: in this case, it is
naturally provided by the non-trivial topology. The fate of the topological contribution to the entanglement after $S_A$ drops to zero at $g\sim0.8$ is discussed at \cite{Note1}.


Increasing $g$ leads to a sucession of new $\log2$
plateaus, indicating that boundary modes are linked with both the
original trivial and topological phases \cite{PhysRevB.83.085426,PhysRevB.86.205119,PhysRevB.94.035144}.
Again, each abrupt change in $S_{A}$ coincides with an anti-crossing
between the $N/2$-th and other states in the upper bands \cite{Note1}. 
Finally, when $g$ approaches the divergence in Fig. \ref{fig:Plotcavityshift} the entropy drops for both phases and the presence of boundary modes is completely washed out. 

\textit{Conclusions and outlook.} $-$ We have investigated the use of the cavity transmission $t_c$ as a topological marker for arbitrary $g$ in a hybrid system, composed of a quantum cavity coupled to a fermionic lattice with distinct topological phases. We consider a topological SSH chain interacting via the dipolar coupling with the cavity, but the framework developed can be applied to the study of other non-trivial topological lattices. We have proven that the measurement of $t_c$ is linked to the topological properties of the unperturbed sample even beyond the small-$g$ regime, when there is a strong hybridization between subsystems. We have considered that the lowest-energy state of the system is occupied, so that the measurement can be performed after the system has thermalized. 

The current state-of-the-art techniques would allow to implement our findings with superconducting qubits interacting with microwave photons \cite{SCCandCav}, where $g$ can reach values comparable to the energy of the qubit \cite{USCsc1, USCsc2, USCsc3, BeyondUSC}. Additionally, the SSH model has also been realized using superconducting qubits \cite{SCsshmodel}. Quantum dots in c-QED architectures have also verified the crucial condition $g>\{\kappa, \gamma\}$ \cite{SCofQDResonator,StrongCouplingEnsslinSQUID,StrongCouplingPetta,StrongCouplingChargeKontos,Strong2DQDEnsslin,2dqdstrongcoupling,GatedQDstrongly,StrongCouplingSpinVandersypen,StrongSpinMonicaPetta,strongspincarbon,strong2spinpetta,StrongSPcouplingSi, StrongTQDEnsslin,StrongTQDEnsslinAbadillo}, though the values of $g$ that are reached would only allow for topological detection in the small-$g$ regime, occupying an edge state \cite{Note1}. In this set-up, aligning the cavity electric field with the axis of the chain would lead to the dipole interaction, as considered in \cite{TQDcavity, TQDcavity2} and implemented in \cite{StrongTQDEnsslin,StrongTQDEnsslinAbadillo} for triple quantum dots.

We believe our work provides a solid basis to study topological systems coupled to quantum cavities, while opening the way to complementary research. Lastly, the employment of the entanglement entropy to study the effect of the interaction upon the fermionic system sets a precedent in the field, since it is usually used to characterize topology in non-interacting systems.

\section*{Acknowledgements}
We thank M. Benito and S. Kohler for fruitful discussions. This work was supported by: Ministerio de Econom\'ia y Competitividad, through Grant MAT2017- 86717-P (CSIC Research Platform PTI-001) and PID2020-117787GB-I00,  Ministerio de Educaci\'on y Formaci\'on Profesional, under the program FPU, with reference FPU17/05297 (B. P.-G.), the Spanish project PGC2018-094792-B-100 (MCIU/AEI/FEDER, EU) (A. G.-L.).

\bibliographystyle{apsrev}
\bibliography{cavitybib}

\begin{thebibliography}{46}
\expandafter\ifx\csname natexlab\endcsname\relax\def\natexlab#1{#1}\fi
\expandafter\ifx\csname bibnamefont\endcsname\relax
  \def\bibnamefont#1{#1}\fi
\expandafter\ifx\csname bibfnamefont\endcsname\relax
  \def\bibfnamefont#1{#1}\fi
\expandafter\ifx\csname citenamefont\endcsname\relax
  \def\citenamefont#1{#1}\fi
\expandafter\ifx\csname url\endcsname\relax
  \def\url#1{\texttt{#1}}\fi
\expandafter\ifx\csname urlprefix\endcsname\relax\def\urlprefix{URL }\fi
\providecommand{\bibinfo}[2]{#2}
\providecommand{\eprint}[2][]{\url{#2}}

\bibitem[{\citenamefont{Walther et~al.}(2006)\citenamefont{Walther, Varcoe,
  Englert, and Becker}}]{AtomicQED}
\bibinfo{author}{\bibfnamefont{H.}~\bibnamefont{Walther}},
  \bibinfo{author}{\bibfnamefont{B.~T.~H.} \bibnamefont{Varcoe}},
  \bibinfo{author}{\bibfnamefont{B.-G.} \bibnamefont{Englert}},
  \bibnamefont{and} \bibinfo{author}{\bibfnamefont{T.}~\bibnamefont{Becker}},
  \bibinfo{journal}{Reports on Progress in Physics}
  \textbf{\bibinfo{volume}{69}}, \bibinfo{pages}{1325} (\bibinfo{year}{2006}),
  \urlprefix\url{https://doi.org/10.1088%2F0034-4885%2F69%2F5%2Fr02}.

\bibitem[{\citenamefont{Cottet et~al.}(2017)\citenamefont{Cottet, C.~Dartiailh,
  M.~Desjardins, Cubaynes, C.~Contamin, Delbecq, J.~Viennot, Doucot, and
  Kontos}}]{CavityQEDAtomToCond}
\bibinfo{author}{\bibfnamefont{A.}~\bibnamefont{Cottet}},
  \bibinfo{author}{\bibfnamefont{M.}~\bibnamefont{C.~Dartiailh}},
  \bibinfo{author}{\bibfnamefont{M.}~\bibnamefont{M.~Desjardins}},
  \bibinfo{author}{\bibfnamefont{T.}~\bibnamefont{Cubaynes}},
  \bibinfo{author}{\bibfnamefont{L.}~\bibnamefont{C.~Contamin}},
  \bibinfo{author}{\bibfnamefont{M.}~\bibnamefont{Delbecq}},
  \bibinfo{author}{\bibfnamefont{J.}~\bibnamefont{J.~Viennot}},
  \bibinfo{author}{\bibfnamefont{B.}~\bibnamefont{Doucot}}, \bibnamefont{and}
  \bibinfo{author}{\bibfnamefont{T.}~\bibnamefont{Kontos}},
  \bibinfo{journal}{J. Phys.: Condens. Matter} \textbf{\bibinfo{volume}{29}}
  (\bibinfo{year}{2017}).

\bibitem[{\citenamefont{Li and Eckstein}(2020)}]{ManipulatingIntertwined}
\bibinfo{author}{\bibfnamefont{J.}~\bibnamefont{Li}} \bibnamefont{and}
  \bibinfo{author}{\bibfnamefont{M.}~\bibnamefont{Eckstein}},
  \bibinfo{journal}{Phys. Rev. Lett.} \textbf{\bibinfo{volume}{125}},
  \bibinfo{pages}{217402} (\bibinfo{year}{2020}),
  \urlprefix\url{https://link.aps.org/doi/10.1103/PhysRevLett.125.217402}.

\bibitem[{\citenamefont{Blais et~al.}(2004)\citenamefont{Blais, Huang,
  Wallraff, Girvin, and Schoelkopf}}]{QEDarchQC}
\bibinfo{author}{\bibfnamefont{A.}~\bibnamefont{Blais}},
  \bibinfo{author}{\bibfnamefont{R.-S.} \bibnamefont{Huang}},
  \bibinfo{author}{\bibfnamefont{A.}~\bibnamefont{Wallraff}},
  \bibinfo{author}{\bibfnamefont{S.~M.} \bibnamefont{Girvin}},
  \bibnamefont{and} \bibinfo{author}{\bibfnamefont{R.~J.}
  \bibnamefont{Schoelkopf}}, \bibinfo{journal}{Phys. Rev. A}
  \textbf{\bibinfo{volume}{69}}, \bibinfo{pages}{062320}
  (\bibinfo{year}{2004}),
  \urlprefix\url{https://link.aps.org/doi/10.1103/PhysRevA.69.062320}.

\bibitem[{\citenamefont{Pellizzari et~al.}(1995)\citenamefont{Pellizzari,
  Gardiner, Cirac, and Zoller}}]{DecQCCavQED}
\bibinfo{author}{\bibfnamefont{T.}~\bibnamefont{Pellizzari}},
  \bibinfo{author}{\bibfnamefont{S.~A.} \bibnamefont{Gardiner}},
  \bibinfo{author}{\bibfnamefont{J.~I.} \bibnamefont{Cirac}}, \bibnamefont{and}
  \bibinfo{author}{\bibfnamefont{P.}~\bibnamefont{Zoller}},
  \bibinfo{journal}{Phys. Rev. Lett.} \textbf{\bibinfo{volume}{75}},
  \bibinfo{pages}{3788} (\bibinfo{year}{1995}).

\bibitem[{\citenamefont{Blais et~al.}(2007)\citenamefont{Blais, Gambetta,
  Wallraff, Schuster, Girvin, Devoret, and Schoelkopf}}]{QInfoCircuitQED}
\bibinfo{author}{\bibfnamefont{A.}~\bibnamefont{Blais}},
  \bibinfo{author}{\bibfnamefont{J.}~\bibnamefont{Gambetta}},
  \bibinfo{author}{\bibfnamefont{A.}~\bibnamefont{Wallraff}},
  \bibinfo{author}{\bibfnamefont{D.~I.} \bibnamefont{Schuster}},
  \bibinfo{author}{\bibfnamefont{S.~M.} \bibnamefont{Girvin}},
  \bibinfo{author}{\bibfnamefont{M.~H.} \bibnamefont{Devoret}},
  \bibnamefont{and} \bibinfo{author}{\bibfnamefont{R.~J.}
  \bibnamefont{Schoelkopf}}, \bibinfo{journal}{Phys. Rev. A}
  \textbf{\bibinfo{volume}{75}}, \bibinfo{pages}{032329}
  (\bibinfo{year}{2007}).

\bibitem[{\citenamefont{Matsukevich and Kuzmich}(2004)}]{StateTransfer}
\bibinfo{author}{\bibfnamefont{D.}~\bibnamefont{Matsukevich}} \bibnamefont{and}
  \bibinfo{author}{\bibfnamefont{A.}~\bibnamefont{Kuzmich}},
  \bibinfo{journal}{Science} \textbf{\bibinfo{volume}{306}},
  \bibinfo{pages}{663} (\bibinfo{year}{2004}).

\bibitem[{\citenamefont{Cirac et~al.}(1997)\citenamefont{Cirac, Zoller, Kimble,
  and Mabuchi}}]{QSTinDistantNodes}
\bibinfo{author}{\bibfnamefont{J.~I.} \bibnamefont{Cirac}},
  \bibinfo{author}{\bibfnamefont{P.}~\bibnamefont{Zoller}},
  \bibinfo{author}{\bibfnamefont{H.~J.} \bibnamefont{Kimble}},
  \bibnamefont{and} \bibinfo{author}{\bibfnamefont{H.}~\bibnamefont{Mabuchi}},
  \bibinfo{journal}{Phys. Rev. Lett.} \textbf{\bibinfo{volume}{78}},
  \bibinfo{pages}{3221} (\bibinfo{year}{1997}),
  \urlprefix\url{https://link.aps.org/doi/10.1103/PhysRevLett.78.3221}.

\bibitem[{\citenamefont{Downing et~al.}(2019)\citenamefont{Downing, Sturges,
  Weick, Stobi\ifmmode~\acute{n}\else \'{n}\fi{}ska, and
  Mart\'{\i}n-Moreno}}]{TopCav}
\bibinfo{author}{\bibfnamefont{C.~A.} \bibnamefont{Downing}},
  \bibinfo{author}{\bibfnamefont{T.~J.} \bibnamefont{Sturges}},
  \bibinfo{author}{\bibfnamefont{G.}~\bibnamefont{Weick}},
  \bibinfo{author}{\bibfnamefont{M.}~\bibnamefont{Stobi\ifmmode~\acute{n}\else
  \'{n}\fi{}ska}}, \bibnamefont{and}
  \bibinfo{author}{\bibfnamefont{L.}~\bibnamefont{Mart\'{\i}n-Moreno}},
  \bibinfo{journal}{Phys. Rev. Lett.} \textbf{\bibinfo{volume}{123}},
  \bibinfo{pages}{217401} (\bibinfo{year}{2019}),
  \urlprefix\url{https://link.aps.org/doi/10.1103/PhysRevLett.123.217401}.

\bibitem[{\citenamefont{Nie and Liu}(2020)}]{TopCav2}
\bibinfo{author}{\bibfnamefont{W.}~\bibnamefont{Nie}} \bibnamefont{and}
  \bibinfo{author}{\bibfnamefont{Y.-x.} \bibnamefont{Liu}},
  \bibinfo{journal}{Phys. Rev. Research} \textbf{\bibinfo{volume}{2}},
  \bibinfo{pages}{012076} (\bibinfo{year}{2020}),
  \urlprefix\url{https://link.aps.org/doi/10.1103/PhysRevResearch.2.012076}.

\bibitem[{\citenamefont{Dartiailh et~al.}(2017)\citenamefont{Dartiailh, Kontos,
  Dou\ifmmode~\mbox{\c{c}}\else \c{c}\fi{}ot, and Cottet}}]{MajoranaDet}
\bibinfo{author}{\bibfnamefont{M.~C.} \bibnamefont{Dartiailh}},
  \bibinfo{author}{\bibfnamefont{T.}~\bibnamefont{Kontos}},
  \bibinfo{author}{\bibfnamefont{B.}~\bibnamefont{Dou\ifmmode~\mbox{\c{c}}\else
  \c{c}\fi{}ot}}, \bibnamefont{and}
  \bibinfo{author}{\bibfnamefont{A.}~\bibnamefont{Cottet}},
  \bibinfo{journal}{Phys. Rev. Lett.} \textbf{\bibinfo{volume}{118}},
  \bibinfo{pages}{126803} (\bibinfo{year}{2017}),
  \urlprefix\url{https://link.aps.org/doi/10.1103/PhysRevLett.118.126803}.

\bibitem[{\citenamefont{Dmytruk et~al.}(2015)\citenamefont{Dmytruk, Trif, and
  Simon}}]{cQEDwithTopSup}
\bibinfo{author}{\bibfnamefont{O.}~\bibnamefont{Dmytruk}},
  \bibinfo{author}{\bibfnamefont{M.}~\bibnamefont{Trif}}, \bibnamefont{and}
  \bibinfo{author}{\bibfnamefont{P.}~\bibnamefont{Simon}},
  \bibinfo{journal}{Phys. Rev. B} \textbf{\bibinfo{volume}{92}},
  \bibinfo{pages}{245432} (\bibinfo{year}{2015}).

\bibitem[{\citenamefont{P\'erez-Gonz\'alez
  et~al.}(2019{\natexlab{a}})\citenamefont{P\'erez-Gonz\'alez, Bello,
  G\'omez-Le\'on, and Platero}}]{mipaper1}
\bibinfo{author}{\bibfnamefont{B.}~\bibnamefont{P\'erez-Gonz\'alez}},
  \bibinfo{author}{\bibfnamefont{M.}~\bibnamefont{Bello}},
  \bibinfo{author}{\bibfnamefont{A.}~\bibnamefont{G\'omez-Le\'on}},
  \bibnamefont{and} \bibinfo{author}{\bibfnamefont{G.}~\bibnamefont{Platero}},
  \bibinfo{journal}{Phys. Rev. B} \textbf{\bibinfo{volume}{99}},
  \bibinfo{pages}{035146} (\bibinfo{year}{2019}{\natexlab{a}}),
  \urlprefix\url{https://link.aps.org/doi/10.1103/PhysRevB.99.035146}.

\bibitem[{\citenamefont{P\'erez-Gonz\'alez
  et~al.}(2019{\natexlab{b}})\citenamefont{P\'erez-Gonz\'alez, Bello, Platero,
  and G\'omez-Le\'on}}]{mipaper2}
\bibinfo{author}{\bibfnamefont{B.}~\bibnamefont{P\'erez-Gonz\'alez}},
  \bibinfo{author}{\bibfnamefont{M.}~\bibnamefont{Bello}},
  \bibinfo{author}{\bibfnamefont{G.}~\bibnamefont{Platero}}, \bibnamefont{and}
  \bibinfo{author}{\bibfnamefont{A.}~\bibnamefont{G\'omez-Le\'on}},
  \bibinfo{journal}{Phys. Rev. Lett.} \textbf{\bibinfo{volume}{123}},
  \bibinfo{pages}{126401} (\bibinfo{year}{2019}{\natexlab{b}}).

\bibitem[{\citenamefont{Bello et~al.}(2016)\citenamefont{Bello, Creffield, and
  Platero}}]{doublonMiguel}
\bibinfo{author}{\bibfnamefont{M.}~\bibnamefont{Bello}},
  \bibinfo{author}{\bibfnamefont{C.}~\bibnamefont{Creffield}},
  \bibnamefont{and} \bibinfo{author}{\bibfnamefont{G.}~\bibnamefont{Platero}},
  \bibinfo{journal}{Scientific Reports} \textbf{\bibinfo{volume}{6}}
  (\bibinfo{year}{2016}).

\bibitem[{\citenamefont{Gardiner and Zoller}(2004)}]{QuantumNoise}
\bibinfo{author}{\bibfnamefont{C.}~\bibnamefont{Gardiner}} \bibnamefont{and}
  \bibinfo{author}{\bibfnamefont{P.}~\bibnamefont{Zoller}},
  \emph{\bibinfo{title}{Quantum Noise: A Handbook of Markovian and
  Non-Markovian Quantum Stochastic Methods with Applications to Quantum
  Optics}} (\bibinfo{publisher}{Springer-Verlag Berlin Heidelberg},
  \bibinfo{year}{2004}).

\bibitem[{Not()}]{Note1}
\bibinfo{note}{See Supplementary Information for further details on the
  analytical calculations regarding input-output theory, photonic Green
  functions, and the effective Hamiltonian, as well as additional numerical
  results, as the solution for $\langle Z \rangle$, the example of topological
  detection occupying the edge state, or the connection between the
  entanglement entropy and the energy spectrum of the system}.

\bibitem[{\citenamefont{Kohler}(2018)}]{PRASigmund}
\bibinfo{author}{\bibfnamefont{S.}~\bibnamefont{Kohler}},
  \bibinfo{journal}{Phys. Rev. A} \textbf{\bibinfo{volume}{98}},
  \bibinfo{pages}{023849} (\bibinfo{year}{2018}),
  \urlprefix\url{https://link.aps.org/doi/10.1103/PhysRevA.98.023849}.

\bibitem[{\citenamefont{Cottet et~al.}(2015)\citenamefont{Cottet, Kontos, and
  Dou\ifmmode~\mbox{\c{c}}\else \c{c}\fi{}ot}}]{ElectronPhotonMesoscopic}
\bibinfo{author}{\bibfnamefont{A.}~\bibnamefont{Cottet}},
  \bibinfo{author}{\bibfnamefont{T.}~\bibnamefont{Kontos}}, \bibnamefont{and}
  \bibinfo{author}{\bibfnamefont{B.}~\bibnamefont{Dou\ifmmode~\mbox{\c{c}}\else
  \c{c}\fi{}ot}}, \bibinfo{journal}{Phys. Rev. B}
  \textbf{\bibinfo{volume}{91}}, \bibinfo{pages}{205417}
  (\bibinfo{year}{2015}),
  \urlprefix\url{https://link.aps.org/doi/10.1103/PhysRevB.91.205417}.

\bibitem[{\citenamefont{Ryu and Hatsugai}(2006)}]{entanglementBerry}
\bibinfo{author}{\bibfnamefont{S.}~\bibnamefont{Ryu}} \bibnamefont{and}
  \bibinfo{author}{\bibfnamefont{Y.}~\bibnamefont{Hatsugai}},
  \bibinfo{journal}{Phys. Rev. B} \textbf{\bibinfo{volume}{73}},
  \bibinfo{pages}{245115} (\bibinfo{year}{2006}),
  \urlprefix\url{https://link.aps.org/doi/10.1103/PhysRevB.73.245115}.

\bibitem[{\citenamefont{Gurarie}(2011)}]{PhysRevB.83.085426}
\bibinfo{author}{\bibfnamefont{V.}~\bibnamefont{Gurarie}},
  \bibinfo{journal}{Phys. Rev. B} \textbf{\bibinfo{volume}{83}},
  \bibinfo{pages}{085426} (\bibinfo{year}{2011}),
  \urlprefix\url{https://link.aps.org/doi/10.1103/PhysRevB.83.085426}.

\bibitem[{\citenamefont{Manmana et~al.}(2012)\citenamefont{Manmana, Essin,
  Noack, and Gurarie}}]{PhysRevB.86.205119}
\bibinfo{author}{\bibfnamefont{S.~R.} \bibnamefont{Manmana}},
  \bibinfo{author}{\bibfnamefont{A.~M.} \bibnamefont{Essin}},
  \bibinfo{author}{\bibfnamefont{R.~M.} \bibnamefont{Noack}}, \bibnamefont{and}
  \bibinfo{author}{\bibfnamefont{V.}~\bibnamefont{Gurarie}},
  \bibinfo{journal}{Phys. Rev. B} \textbf{\bibinfo{volume}{86}},
  \bibinfo{pages}{205119} (\bibinfo{year}{2012}),
  \urlprefix\url{https://link.aps.org/doi/10.1103/PhysRevB.86.205119}.

\bibitem[{\citenamefont{G\'omez-Le\'on}(2016)}]{PhysRevB.94.035144}
\bibinfo{author}{\bibfnamefont{A.}~\bibnamefont{G\'omez-Le\'on}},
  \bibinfo{journal}{Phys. Rev. B} \textbf{\bibinfo{volume}{94}},
  \bibinfo{pages}{035144} (\bibinfo{year}{2016}),
  \urlprefix\url{https://link.aps.org/doi/10.1103/PhysRevB.94.035144}.

\bibitem[{\citenamefont{Gu et~al.}(2017)\citenamefont{Gu, Kockum, Miranowicz,
  xi~Liu, and Nori}}]{SCCandCav}
\bibinfo{author}{\bibfnamefont{X.}~\bibnamefont{Gu}},
  \bibinfo{author}{\bibfnamefont{A.~F.} \bibnamefont{Kockum}},
  \bibinfo{author}{\bibfnamefont{A.}~\bibnamefont{Miranowicz}},
  \bibinfo{author}{\bibfnamefont{Y.}~\bibnamefont{xi~Liu}}, \bibnamefont{and}
  \bibinfo{author}{\bibfnamefont{F.}~\bibnamefont{Nori}},
  \bibinfo{journal}{Physics Reports} \textbf{\bibinfo{volume}{718-719}},
  \bibinfo{pages}{1} (\bibinfo{year}{2017}), ISSN \bibinfo{issn}{0370-1573},
  \bibinfo{note}{microwave photonics with superconducting quantum circuits},
  \urlprefix\url{https://www.sciencedirect.com/science/article/pii/S0370157317303290}.

\bibitem[{\citenamefont{Baust et~al.}(2016)\citenamefont{Baust, Hoffmann,
  Haeberlein, Schwarz, Eder, Goetz, Wulschner, Xie, Zhong, Quijandr\'{\i}a
  et~al.}}]{USCsc1}
\bibinfo{author}{\bibfnamefont{A.}~\bibnamefont{Baust}},
  \bibinfo{author}{\bibfnamefont{E.}~\bibnamefont{Hoffmann}},
  \bibinfo{author}{\bibfnamefont{M.}~\bibnamefont{Haeberlein}},
  \bibinfo{author}{\bibfnamefont{M.~J.} \bibnamefont{Schwarz}},
  \bibinfo{author}{\bibfnamefont{P.}~\bibnamefont{Eder}},
  \bibinfo{author}{\bibfnamefont{J.}~\bibnamefont{Goetz}},
  \bibinfo{author}{\bibfnamefont{F.}~\bibnamefont{Wulschner}},
  \bibinfo{author}{\bibfnamefont{E.}~\bibnamefont{Xie}},
  \bibinfo{author}{\bibfnamefont{L.}~\bibnamefont{Zhong}},
  \bibinfo{author}{\bibfnamefont{F.}~\bibnamefont{Quijandr\'{\i}a}},
  \bibnamefont{et~al.}, \bibinfo{journal}{Phys. Rev. B}
  \textbf{\bibinfo{volume}{93}}, \bibinfo{pages}{214501}
  (\bibinfo{year}{2016}),
  \urlprefix\url{https://link.aps.org/doi/10.1103/PhysRevB.93.214501}.

\bibitem[{\citenamefont{Forn-D\'{\i}az
  et~al.}(2010)\citenamefont{Forn-D\'{\i}az, Lisenfeld, Marcos,
  Garc\'{\i}a-Ripoll, Solano, Harmans, and Mooij}}]{USCsc2}
\bibinfo{author}{\bibfnamefont{P.}~\bibnamefont{Forn-D\'{\i}az}},
  \bibinfo{author}{\bibfnamefont{J.}~\bibnamefont{Lisenfeld}},
  \bibinfo{author}{\bibfnamefont{D.}~\bibnamefont{Marcos}},
  \bibinfo{author}{\bibfnamefont{J.~J.} \bibnamefont{Garc\'{\i}a-Ripoll}},
  \bibinfo{author}{\bibfnamefont{E.}~\bibnamefont{Solano}},
  \bibinfo{author}{\bibfnamefont{C.~J. P.~M.} \bibnamefont{Harmans}},
  \bibnamefont{and} \bibinfo{author}{\bibfnamefont{J.~E.} \bibnamefont{Mooij}},
  \bibinfo{journal}{Phys. Rev. Lett.} \textbf{\bibinfo{volume}{105}},
  \bibinfo{pages}{237001} (\bibinfo{year}{2010}),
  \urlprefix\url{https://link.aps.org/doi/10.1103/PhysRevLett.105.237001}.

\bibitem[{\citenamefont{Niemczyk et~al.}(2010)\citenamefont{Niemczyk, Deppe,
  Huebl, Menzel, Hocke, Schwarz, Garcia-Ripoll, Zueco, H{\"u}mmer, Solano
  et~al.}}]{USCsc3}
\bibinfo{author}{\bibfnamefont{T.}~\bibnamefont{Niemczyk}},
  \bibinfo{author}{\bibfnamefont{F.}~\bibnamefont{Deppe}},
  \bibinfo{author}{\bibfnamefont{H.}~\bibnamefont{Huebl}},
  \bibinfo{author}{\bibfnamefont{E.~P.} \bibnamefont{Menzel}},
  \bibinfo{author}{\bibfnamefont{F.}~\bibnamefont{Hocke}},
  \bibinfo{author}{\bibfnamefont{M.~J.} \bibnamefont{Schwarz}},
  \bibinfo{author}{\bibfnamefont{J.~J.} \bibnamefont{Garcia-Ripoll}},
  \bibinfo{author}{\bibfnamefont{D.}~\bibnamefont{Zueco}},
  \bibinfo{author}{\bibfnamefont{T.}~\bibnamefont{H{\"u}mmer}},
  \bibinfo{author}{\bibfnamefont{E.}~\bibnamefont{Solano}},
  \bibnamefont{et~al.}, \bibinfo{journal}{Nature Physics}
  \textbf{\bibinfo{volume}{6}}, \bibinfo{pages}{772} (\bibinfo{year}{2010}),
  ISSN \bibinfo{issn}{1745-2481},
  \urlprefix\url{https://doi.org/10.1038/nphys1730}.

\bibitem[{\citenamefont{Yoshihara et~al.}(2017)\citenamefont{Yoshihara, Fuse,
  Ashhab, Kakuyanagi, Saito, and Semba}}]{BeyondUSC}
\bibinfo{author}{\bibfnamefont{F.}~\bibnamefont{Yoshihara}},
  \bibinfo{author}{\bibfnamefont{T.}~\bibnamefont{Fuse}},
  \bibinfo{author}{\bibfnamefont{S.}~\bibnamefont{Ashhab}},
  \bibinfo{author}{\bibfnamefont{K.}~\bibnamefont{Kakuyanagi}},
  \bibinfo{author}{\bibfnamefont{S.}~\bibnamefont{Saito}}, \bibnamefont{and}
  \bibinfo{author}{\bibfnamefont{K.}~\bibnamefont{Semba}},
  \bibinfo{journal}{Nature Physics} \textbf{\bibinfo{volume}{13}},
  \bibinfo{pages}{44} (\bibinfo{year}{2017}), ISSN \bibinfo{issn}{1745-2481},
  \urlprefix\url{https://doi.org/10.1038/nphys3906}.

\bibitem[{\citenamefont{Besedin et~al.}(2021)\citenamefont{Besedin, Gorlach,
  Abramov, Tsitsilin, Moskalenko, Dobronosova, Moskalev, Matanin, Smirnov,
  Rodionov et~al.}}]{SCsshmodel}
\bibinfo{author}{\bibfnamefont{I.~S.} \bibnamefont{Besedin}},
  \bibinfo{author}{\bibfnamefont{M.~A.} \bibnamefont{Gorlach}},
  \bibinfo{author}{\bibfnamefont{N.~N.} \bibnamefont{Abramov}},
  \bibinfo{author}{\bibfnamefont{I.}~\bibnamefont{Tsitsilin}},
  \bibinfo{author}{\bibfnamefont{I.~N.} \bibnamefont{Moskalenko}},
  \bibinfo{author}{\bibfnamefont{A.~A.} \bibnamefont{Dobronosova}},
  \bibinfo{author}{\bibfnamefont{D.~O.} \bibnamefont{Moskalev}},
  \bibinfo{author}{\bibfnamefont{A.~R.} \bibnamefont{Matanin}},
  \bibinfo{author}{\bibfnamefont{N.~S.} \bibnamefont{Smirnov}},
  \bibinfo{author}{\bibfnamefont{I.~A.} \bibnamefont{Rodionov}},
  \bibnamefont{et~al.}, \bibinfo{journal}{Phys. Rev. B}
  \textbf{\bibinfo{volume}{103}}, \bibinfo{pages}{224520}
  (\bibinfo{year}{2021}),
  \urlprefix\url{https://link.aps.org/doi/10.1103/PhysRevB.103.224520}.

\bibitem[{\citenamefont{Stockklauser
  et~al.}(2017{\natexlab{a}})\citenamefont{Stockklauser, Scarlino, Koski,
  Gasparinetti, Andersen, Reichl, Wegscheider, Ihn, Ensslin, and
  Wallraff}}]{SCofQDResonator}
\bibinfo{author}{\bibfnamefont{A.}~\bibnamefont{Stockklauser}},
  \bibinfo{author}{\bibfnamefont{P.}~\bibnamefont{Scarlino}},
  \bibinfo{author}{\bibfnamefont{J.~V.} \bibnamefont{Koski}},
  \bibinfo{author}{\bibfnamefont{S.}~\bibnamefont{Gasparinetti}},
  \bibinfo{author}{\bibfnamefont{C.~K.} \bibnamefont{Andersen}},
  \bibinfo{author}{\bibfnamefont{C.}~\bibnamefont{Reichl}},
  \bibinfo{author}{\bibfnamefont{W.}~\bibnamefont{Wegscheider}},
  \bibinfo{author}{\bibfnamefont{T.}~\bibnamefont{Ihn}},
  \bibinfo{author}{\bibfnamefont{K.}~\bibnamefont{Ensslin}}, \bibnamefont{and}
  \bibinfo{author}{\bibfnamefont{A.}~\bibnamefont{Wallraff}},
  \bibinfo{journal}{Phys. Rev. X} \textbf{\bibinfo{volume}{7}},
  \bibinfo{pages}{011030} (\bibinfo{year}{2017}{\natexlab{a}}),
  \urlprefix\url{https://link.aps.org/doi/10.1103/PhysRevX.7.011030}.

\bibitem[{\citenamefont{Stockklauser
  et~al.}(2017{\natexlab{b}})\citenamefont{Stockklauser, Scarlino, Koski,
  Gasparinetti, Andersen, Reichl, Wegscheider, Ihn, Ensslin, and
  Wallraff}}]{StrongCouplingEnsslinSQUID}
\bibinfo{author}{\bibfnamefont{A.}~\bibnamefont{Stockklauser}},
  \bibinfo{author}{\bibfnamefont{P.}~\bibnamefont{Scarlino}},
  \bibinfo{author}{\bibfnamefont{J.~V.} \bibnamefont{Koski}},
  \bibinfo{author}{\bibfnamefont{S.}~\bibnamefont{Gasparinetti}},
  \bibinfo{author}{\bibfnamefont{C.~K.} \bibnamefont{Andersen}},
  \bibinfo{author}{\bibfnamefont{C.}~\bibnamefont{Reichl}},
  \bibinfo{author}{\bibfnamefont{W.}~\bibnamefont{Wegscheider}},
  \bibinfo{author}{\bibfnamefont{T.}~\bibnamefont{Ihn}},
  \bibinfo{author}{\bibfnamefont{K.}~\bibnamefont{Ensslin}}, \bibnamefont{and}
  \bibinfo{author}{\bibfnamefont{A.}~\bibnamefont{Wallraff}},
  \bibinfo{journal}{Phys. Rev. X} \textbf{\bibinfo{volume}{7}},
  \bibinfo{pages}{011030} (\bibinfo{year}{2017}{\natexlab{b}}),
  \urlprefix\url{https://link.aps.org/doi/10.1103/PhysRevX.7.011030}.

\bibitem[{\citenamefont{Mi et~al.}(2017)\citenamefont{Mi, Cady, Zajac, Deelman,
  and Petta}}]{StrongCouplingPetta}
\bibinfo{author}{\bibfnamefont{X.}~\bibnamefont{Mi}},
  \bibinfo{author}{\bibfnamefont{J.~M.} \bibnamefont{Cady}},
  \bibinfo{author}{\bibfnamefont{D.~M.} \bibnamefont{Zajac}},
  \bibinfo{author}{\bibfnamefont{P.~W.} \bibnamefont{Deelman}},
  \bibnamefont{and} \bibinfo{author}{\bibfnamefont{J.~R.} \bibnamefont{Petta}},
  \bibinfo{journal}{Science} \textbf{\bibinfo{volume}{355}}
  (\bibinfo{year}{2017}).

\bibitem[{\citenamefont{Bruhat et~al.}(2018)\citenamefont{Bruhat, Cubaynes,
  Viennot, Dartiailh, Desjardins, Cottet, and
  Kontos}}]{StrongCouplingChargeKontos}
\bibinfo{author}{\bibfnamefont{L.~E.} \bibnamefont{Bruhat}},
  \bibinfo{author}{\bibfnamefont{T.}~\bibnamefont{Cubaynes}},
  \bibinfo{author}{\bibfnamefont{J.~J.} \bibnamefont{Viennot}},
  \bibinfo{author}{\bibfnamefont{M.~C.} \bibnamefont{Dartiailh}},
  \bibinfo{author}{\bibfnamefont{M.~M.} \bibnamefont{Desjardins}},
  \bibinfo{author}{\bibfnamefont{A.}~\bibnamefont{Cottet}}, \bibnamefont{and}
  \bibinfo{author}{\bibfnamefont{T.}~\bibnamefont{Kontos}},
  \bibinfo{journal}{Phys. Rev. B} \textbf{\bibinfo{volume}{98}},
  \bibinfo{pages}{155313} (\bibinfo{year}{2018}),
  \urlprefix\url{https://link.aps.org/doi/10.1103/PhysRevB.98.155313}.

\bibitem[{\citenamefont{van Woerkom et~al.}(2018)\citenamefont{van Woerkom,
  Scarlino, Ungerer, M\"uller, Koski, Landig, Reichl, Wegscheider, Ihn, Ensslin
  et~al.}}]{Strong2DQDEnsslin}
\bibinfo{author}{\bibfnamefont{D.~J.} \bibnamefont{van Woerkom}},
  \bibinfo{author}{\bibfnamefont{P.}~\bibnamefont{Scarlino}},
  \bibinfo{author}{\bibfnamefont{J.~H.} \bibnamefont{Ungerer}},
  \bibinfo{author}{\bibfnamefont{C.}~\bibnamefont{M\"uller}},
  \bibinfo{author}{\bibfnamefont{J.~V.} \bibnamefont{Koski}},
  \bibinfo{author}{\bibfnamefont{A.~J.} \bibnamefont{Landig}},
  \bibinfo{author}{\bibfnamefont{C.}~\bibnamefont{Reichl}},
  \bibinfo{author}{\bibfnamefont{W.}~\bibnamefont{Wegscheider}},
  \bibinfo{author}{\bibfnamefont{T.}~\bibnamefont{Ihn}},
  \bibinfo{author}{\bibfnamefont{K.}~\bibnamefont{Ensslin}},
  \bibnamefont{et~al.}, \bibinfo{journal}{Phys. Rev. X}
  \textbf{\bibinfo{volume}{8}}, \bibinfo{pages}{041018} (\bibinfo{year}{2018}),
  \urlprefix\url{https://link.aps.org/doi/10.1103/PhysRevX.8.041018}.

\bibitem[{\citenamefont{Wang et~al.}(2021)\citenamefont{Wang, Lin, Li, Gu,
  Chen, Guo, Jiang, Hu, Cao, and Guo}}]{2dqdstrongcoupling}
\bibinfo{author}{\bibfnamefont{B.}~\bibnamefont{Wang}},
  \bibinfo{author}{\bibfnamefont{T.}~\bibnamefont{Lin}},
  \bibinfo{author}{\bibfnamefont{H.}~\bibnamefont{Li}},
  \bibinfo{author}{\bibfnamefont{S.}~\bibnamefont{Gu}},
  \bibinfo{author}{\bibfnamefont{M.}~\bibnamefont{Chen}},
  \bibinfo{author}{\bibfnamefont{G.}~\bibnamefont{Guo}},
  \bibinfo{author}{\bibfnamefont{H.}~\bibnamefont{Jiang}},
  \bibinfo{author}{\bibfnamefont{X.}~\bibnamefont{Hu}},
  \bibinfo{author}{\bibfnamefont{G.}~\bibnamefont{Cao}}, \bibnamefont{and}
  \bibinfo{author}{\bibfnamefont{G.}~\bibnamefont{Guo}},
  \bibinfo{journal}{Science Bulletin} \textbf{\bibinfo{volume}{66}},
  \bibinfo{pages}{332} (\bibinfo{year}{2021}), ISSN \bibinfo{issn}{2095-9273},
  \urlprefix\url{https://www.sciencedirect.com/science/article/pii/S2095927320306587}.

\bibitem[{\citenamefont{Najer et~al.}(2019)\citenamefont{Najer, S\"oller,
  Sekatski, Dolique, L\"obl, Riedel, Schott, Starosielec, Valentin, Wieck
  et~al.}}]{GatedQDstrongly}
\bibinfo{author}{\bibfnamefont{D.}~\bibnamefont{Najer}},
  \bibinfo{author}{\bibfnamefont{I.}~\bibnamefont{S\"oller}},
  \bibinfo{author}{\bibfnamefont{P.}~\bibnamefont{Sekatski}},
  \bibinfo{author}{\bibfnamefont{V.}~\bibnamefont{Dolique}},
  \bibinfo{author}{\bibfnamefont{M.~C.} \bibnamefont{L\"obl}},
  \bibinfo{author}{\bibfnamefont{D.}~\bibnamefont{Riedel}},
  \bibinfo{author}{\bibfnamefont{R.}~\bibnamefont{Schott}},
  \bibinfo{author}{\bibfnamefont{S.}~\bibnamefont{Starosielec}},
  \bibinfo{author}{\bibfnamefont{S.~R.} \bibnamefont{Valentin}},
  \bibinfo{author}{\bibfnamefont{A.}~\bibnamefont{Wieck}},
  \bibnamefont{et~al.}, \bibinfo{journal}{Nature} \textbf{\bibinfo{volume}{5}},
  \bibinfo{pages}{622} (\bibinfo{year}{2019}).

\bibitem[{\citenamefont{Samkharadze
  et~al.}(2018{\natexlab{a}})\citenamefont{Samkharadze, Zheng, Kalhor, Brousse,
  Sammak, Mendes, Blais, Scappucci, and
  Vadersypen}}]{StrongCouplingSpinVandersypen}
\bibinfo{author}{\bibfnamefont{N.}~\bibnamefont{Samkharadze}},
  \bibinfo{author}{\bibfnamefont{G.}~\bibnamefont{Zheng}},
  \bibinfo{author}{\bibfnamefont{N.}~\bibnamefont{Kalhor}},
  \bibinfo{author}{\bibfnamefont{D.}~\bibnamefont{Brousse}},
  \bibinfo{author}{\bibfnamefont{A.}~\bibnamefont{Sammak}},
  \bibinfo{author}{\bibfnamefont{U.~C.} \bibnamefont{Mendes}},
  \bibinfo{author}{\bibfnamefont{A.}~\bibnamefont{Blais}},
  \bibinfo{author}{\bibfnamefont{G.}~\bibnamefont{Scappucci}},
  \bibnamefont{and} \bibinfo{author}{\bibfnamefont{L.~M.~K.}
  \bibnamefont{Vadersypen}}, \bibinfo{journal}{Science}
  \textbf{\bibinfo{volume}{359}} (\bibinfo{year}{2018}{\natexlab{a}}).

\bibitem[{\citenamefont{Mi et~al.}(2018)\citenamefont{Mi, Benito, Putz, Zajac,
  Taylor, Burkard, and Petta}}]{StrongSpinMonicaPetta}
\bibinfo{author}{\bibfnamefont{X.}~\bibnamefont{Mi}},
  \bibinfo{author}{\bibfnamefont{M.}~\bibnamefont{Benito}},
  \bibinfo{author}{\bibfnamefont{S.}~\bibnamefont{Putz}},
  \bibinfo{author}{\bibfnamefont{D.~M.} \bibnamefont{Zajac}},
  \bibinfo{author}{\bibfnamefont{J.~M.} \bibnamefont{Taylor}},
  \bibinfo{author}{\bibfnamefont{G.}~\bibnamefont{Burkard}}, \bibnamefont{and}
  \bibinfo{author}{\bibfnamefont{J.~R.} \bibnamefont{Petta}},
  \bibinfo{journal}{Nature} \textbf{\bibinfo{volume}{555}},
  \bibinfo{pages}{590} (\bibinfo{year}{2018}).

\bibitem[{\citenamefont{Cubaynes et~al.}(2019)\citenamefont{Cubaynes, Delbecq,
  Dartiailh, Assouly, Desjardins, Contamin, Bruhat, Leghtas, Mallet, Cottet
  et~al.}}]{strongspincarbon}
\bibinfo{author}{\bibfnamefont{T.}~\bibnamefont{Cubaynes}},
  \bibinfo{author}{\bibfnamefont{M.~R.} \bibnamefont{Delbecq}},
  \bibinfo{author}{\bibfnamefont{M.~C.} \bibnamefont{Dartiailh}},
  \bibinfo{author}{\bibfnamefont{R.}~\bibnamefont{Assouly}},
  \bibinfo{author}{\bibfnamefont{M.~M.} \bibnamefont{Desjardins}},
  \bibinfo{author}{\bibfnamefont{L.~C.} \bibnamefont{Contamin}},
  \bibinfo{author}{\bibfnamefont{L.~E.} \bibnamefont{Bruhat}},
  \bibinfo{author}{\bibfnamefont{Z.}~\bibnamefont{Leghtas}},
  \bibinfo{author}{\bibfnamefont{F.}~\bibnamefont{Mallet}},
  \bibinfo{author}{\bibfnamefont{A.}~\bibnamefont{Cottet}},
  \bibnamefont{et~al.}, \bibinfo{journal}{npj Quantum Information}
  \textbf{\bibinfo{volume}{5}} (\bibinfo{year}{2019}).

\bibitem[{\citenamefont{Borjans et~al.}(2020)\citenamefont{Borjans, Mi,
  Gullans, and Petta}}]{strong2spinpetta}
\bibinfo{author}{\bibfnamefont{X.~G.} \bibnamefont{Borjans},
  \bibfnamefont{F.~Croot}},
  \bibinfo{author}{\bibfnamefont{X.}~\bibnamefont{Mi}},
  \bibinfo{author}{\bibfnamefont{M.~J.} \bibnamefont{Gullans}},
  \bibnamefont{and} \bibinfo{author}{\bibfnamefont{J.~R.} \bibnamefont{Petta}},
  \bibinfo{journal}{Nature} \textbf{\bibinfo{volume}{577}},
  \bibinfo{pages}{195} (\bibinfo{year}{2020}).

\bibitem[{\citenamefont{Samkharadze
  et~al.}(2018{\natexlab{b}})\citenamefont{Samkharadze, Zheng, Kalhor, Brousse,
  Sammak, Mendes, Blais, Scappucci, and Vandersypen}}]{StrongSPcouplingSi}
\bibinfo{author}{\bibfnamefont{N.}~\bibnamefont{Samkharadze}},
  \bibinfo{author}{\bibfnamefont{G.}~\bibnamefont{Zheng}},
  \bibinfo{author}{\bibfnamefont{N.}~\bibnamefont{Kalhor}},
  \bibinfo{author}{\bibfnamefont{D.}~\bibnamefont{Brousse}},
  \bibinfo{author}{\bibfnamefont{A.}~\bibnamefont{Sammak}},
  \bibinfo{author}{\bibfnamefont{U.~C.} \bibnamefont{Mendes}},
  \bibinfo{author}{\bibfnamefont{A.}~\bibnamefont{Blais}},
  \bibinfo{author}{\bibfnamefont{G.}~\bibnamefont{Scappucci}},
  \bibnamefont{and} \bibinfo{author}{\bibfnamefont{L.~M.~K.}
  \bibnamefont{Vandersypen}}, \bibinfo{journal}{Science}
  \textbf{\bibinfo{volume}{359}}, \bibinfo{pages}{1123}
  (\bibinfo{year}{2018}{\natexlab{b}}).

\bibitem[{\citenamefont{Landig et~al.}(2018)\citenamefont{Landig, Koski,
  Scalino, Mendes, Blais, Reichl, Wegscheider, Wallraff, Ensslin, and
  Ihn}}]{StrongTQDEnsslin}
\bibinfo{author}{\bibfnamefont{A.~J.} \bibnamefont{Landig}},
  \bibinfo{author}{\bibfnamefont{J.~V.} \bibnamefont{Koski}},
  \bibinfo{author}{\bibfnamefont{P.}~\bibnamefont{Scalino}},
  \bibinfo{author}{\bibfnamefont{U.~C.} \bibnamefont{Mendes}},
  \bibinfo{author}{\bibfnamefont{A.}~\bibnamefont{Blais}},
  \bibinfo{author}{\bibfnamefont{C.}~\bibnamefont{Reichl}},
  \bibinfo{author}{\bibfnamefont{W.}~\bibnamefont{Wegscheider}},
  \bibinfo{author}{\bibfnamefont{A.}~\bibnamefont{Wallraff}},
  \bibinfo{author}{\bibfnamefont{K.}~\bibnamefont{Ensslin}}, \bibnamefont{and}
  \bibinfo{author}{\bibfnamefont{T.}~\bibnamefont{Ihn}},
  \bibinfo{journal}{Nature} \textbf{\bibinfo{volume}{560}},
  \bibinfo{pages}{179} (\bibinfo{year}{2018}).

\bibitem[{\citenamefont{Landig et~al.}(2019)\citenamefont{Landig, Koski,
  Scarlino, M\"uller, Abadillo-Uriel, Kratochwil, Reichl, Wegscheider,
  Coppersmith, Friesen et~al.}}]{StrongTQDEnsslinAbadillo}
\bibinfo{author}{\bibfnamefont{A.~J.} \bibnamefont{Landig}},
  \bibinfo{author}{\bibfnamefont{J.~V.} \bibnamefont{Koski}},
  \bibinfo{author}{\bibfnamefont{P.}~\bibnamefont{Scarlino}},
  \bibinfo{author}{\bibfnamefont{C.}~\bibnamefont{M\"uller}},
  \bibinfo{author}{\bibfnamefont{J.~C.} \bibnamefont{Abadillo-Uriel}},
  \bibinfo{author}{\bibfnamefont{B.}~\bibnamefont{Kratochwil}},
  \bibinfo{author}{\bibfnamefont{C.}~\bibnamefont{Reichl}},
  \bibinfo{author}{\bibfnamefont{W.}~\bibnamefont{Wegscheider}},
  \bibinfo{author}{\bibfnamefont{S.~N.} \bibnamefont{Coppersmith}},
  \bibinfo{author}{\bibfnamefont{M.}~\bibnamefont{Friesen}},
  \bibnamefont{et~al.}, \bibinfo{journal}{Nature Communications}
  \textbf{\bibinfo{volume}{10}} (\bibinfo{year}{2019}).

\bibitem[{\citenamefont{Russ et~al.}(2016)\citenamefont{Russ, Ginzel, and
  Burkard}}]{TQDcavity}
\bibinfo{author}{\bibfnamefont{M.}~\bibnamefont{Russ}},
  \bibinfo{author}{\bibfnamefont{F.}~\bibnamefont{Ginzel}}, \bibnamefont{and}
  \bibinfo{author}{\bibfnamefont{G.}~\bibnamefont{Burkard}},
  \bibinfo{journal}{Phys. Rev. B} \textbf{\bibinfo{volume}{94}},
  \bibinfo{pages}{165411} (\bibinfo{year}{2016}),
  \urlprefix\url{https://link.aps.org/doi/10.1103/PhysRevB.94.165411}.

\bibitem[{\citenamefont{Srinivasa et~al.}(2016)\citenamefont{Srinivasa, Taylor,
  and Tahan}}]{TQDcavity2}
\bibinfo{author}{\bibfnamefont{V.}~\bibnamefont{Srinivasa}},
  \bibinfo{author}{\bibfnamefont{J.~M.} \bibnamefont{Taylor}},
  \bibnamefont{and} \bibinfo{author}{\bibfnamefont{C.}~\bibnamefont{Tahan}},
  \bibinfo{journal}{Phys. Rev. B} \textbf{\bibinfo{volume}{94}},
  \bibinfo{pages}{205421} (\bibinfo{year}{2016}),
  \urlprefix\url{https://link.aps.org/doi/10.1103/PhysRevB.94.205421}.

\bibitem[{\citenamefont{Walls and Milburn}(2008)}]{MilburnWallsQO}
\bibinfo{author}{\bibfnamefont{D.~F.} \bibnamefont{Walls}} \bibnamefont{and}
  \bibinfo{author}{\bibfnamefont{G.~J.} \bibnamefont{Milburn}},
  \emph{\bibinfo{title}{Quantum Optics}} (\bibinfo{publisher}{Springer},
  \bibinfo{year}{2008}).

\end{thebibliography}

\appendix
\section{\label{sec:Input-output-formalism}Input-output formalism}

\subsubsection*{Using the initial Hamiltonian} Consider the initial Hamiltonian describing a transmission line with
two ports, coupled to the topological system via the cavity photons.
The total Hamiltonian is:

\begin{equation}
H=\Omega d^{\dagger}d+H_{S}+g\left(d^{\dagger}+d\right)Z+H_{B},
\end{equation}

with $H_{B}$ being the Hamiltonian for the electromagnetic field
in the transmission line and its coupling to the cavity, 

\begin{eqnarray}
H_{B} & = & \sum_{l=1,2}\int_{-\infty}^{\infty}\omega b_{l}^{\dagger}\left(\omega\right)b_{l}\left(\omega\right)d\omega \nonumber \\
&  & +i\sum_{l=1,2}\int_{-\infty}^{\infty}\left[\mu_{l}\left(\omega\right)b_{l}^{\dagger}\left(\omega\right)d \right. \nonumber \\
& & \hspace{2.9cm} \left. -\mu_{l}\left(\omega\right)^{\ast}d^{\dagger}b_{l}\left(\omega\right) \right] d\omega 
\end{eqnarray}

where $l=1,2$ represent the left/right sides of the cavity, $b_{l}\left(\omega\right)$
is the destruction operator for a photon with energy $\omega$ at
side $l$ of the cavity and $\mu_{l}(\omega)$ represents the coupling
between the cavity and the outside modes. In this work,  we will consider the Markov approximation, which gives $\mu_{l}\left(\omega\right) =  \sqrt{\kappa_{l}/2\pi}$.\\

Then, the first step is to derive the equation
of motion (EoM) for the transmission line photons:

\begin{equation}
\frac{d}{dt}b_{l}\left(\nu,t\right)=-i\nu b_{l}\left(\nu,t\right)+\sqrt{\frac{\kappa_{l}}{2\pi}}d\left(t\right)
\end{equation}

which can be formally integrated to yield

\begin{eqnarray}
b_{l}\left(\nu,t\right) & = &  b_{l}\left(\nu,t_{0}\right)e^{-i\nu\left(t-t_{0}\right)} \nonumber \\ 
&  & +  \sqrt{\frac{\kappa_{l}}{2\pi}}\int_{t_{0}}^{t}dt^{\prime}e^{-i\nu\left(t-t^{\prime}\right)}d\left(t^{\prime}\right) 
\label{eq:solforb}
\end{eqnarray}

where $t_{0}<t$ represents the initial condition. Inserting the previous
expression into the EoM for the cavity photons, we get

\begin{eqnarray}
\frac{d}{dt}d\left(t\right) & = & -i\Omega d\left(t\right)-igZ(t)-\sum_{l=1,2}\sqrt{\frac{\kappa_{l}}{2\pi}}\int_{-\infty}^{\infty}d\nu b_{l}\left(\nu,t\right)\nonumber \\
& = & -i\left(\Omega-i\frac{\kappa}{2}\right)d\left(t\right)-igZ\left(t\right) \nonumber \\
& & \hspace{60pt} -\sum_{l=1,2}\sqrt{\kappa_{l}}b_{in,l}\left(t\right)
\label{eq:dadt}
\end{eqnarray}

where $\kappa=\kappa_{1}+\kappa_{2}$, and we have defined an input
field,

\begin{eqnarray}
b_{in,l}\left(t\right) & = & \frac{1}{\sqrt{2\pi}}\int_{-\infty}^{\infty}b_{l}\left(\nu,t_{0}\right)e^{-i\nu\left(t-t_{0}\right)}d\nu.
\end{eqnarray}

Similarly, the solution for $b_{l}\left(\nu,t\right)$ in (\ref{eq:solforb})
can also be obtained in terms of a final condition $t_{1}>t$, which
let us define an output field $b_{\mathrm{out},l}\left(t\right)$, fulfilling

\begin{equation}
b_{\mathrm{out},l}\left(t\right)=b_{\mathrm{in},l}\left(t\right)+\sqrt{\kappa_{l}}d\left(t\right)\label{eq:inandout}
\end{equation}

One has to solve the EoM for $Z\left(t\right)$ as well. We consider
the basis of eigenstates of the fermionic Hamiltonian

\begin{equation}
H_{S}=\sum_{\alpha=1}^{N}E_{\alpha}X^{\alpha,\alpha},\ Z=\sum_{\vec{\alpha}}Z_{\vec{\alpha}}X^{\vec{\alpha}}
\end{equation}

with $X^{\vec{\alpha}} = X^{\alpha_1,\alpha_2} = |\alpha_1\rangle \langle \alpha_2 |$, and calculate the equation of motion for an arbitrary Hubbard operator
$X^{\vec{\alpha}}$

\begin{eqnarray}
\partial_{t}\tilde{X}^{\vec{\alpha}}(t) & = &  i\left(\tilde{E}_{\vec{\alpha}} - i\frac{\gamma}{2}\right)\tilde{X}^{\vec{\alpha}}(t) \nonumber \\
& & + ig\left(d^{\dagger}(t) + d(t) \right)\sum_{\beta}\left(\tilde{Z}_{\beta,\alpha_{1}}\tilde{X}^{\beta,\alpha_{2}}(t) \right. \nonumber \\
& & \hspace{70pt} \left. -\tilde{Z}_{\alpha_{2},\beta}\tilde{X}^{\alpha_{1},\beta}(t)\right),
\label{eomX}
\end{eqnarray}

where $E_{\vec{\alpha}} = E_{\alpha_1} - E_{\alpha_2}$ and we have also included the phenomenological spectral broadening factor $\gamma/2$.
The product $d^{\left(\dagger\right)}\left(t\right)X\left(t\right)$
can be decomposed as 

\begin{equation}
d^{\left(\dagger\right)}\left(t\right)X\left(t\right)\approx\langle X\rangle d^{\left(\dagger\right)}\left(t\right) + \langle d^{\left(\dagger\right)}\rangle X(t), 
\end{equation}

where we are neglecting any terms accounting for correlation between operators, which is a valid approach in the small-$g$ regime. Under this condition, one can safely calculate $\langle X \rangle $ and $\langle d^{(\dagger)} \rangle$ using the corresponding unperturbed Hamiltonians, $H_S$ and $\Omega d^\dagger d$, respectively. Then, one can easily see that $\langle d^{(\dagger)} \rangle = 0$ , and hence the solution for the EoM in (\ref{eomX}) in frequency space reads 

\begin{equation}
X^{\vec{\alpha}}\left(\omega\right)\simeq gd\left(\omega\right)\frac{\sum_{\beta}\left(Z_{\alpha_{2},\beta}\langle X^{\alpha_{1},\beta}\rangle-Z_{\beta,\alpha_{1}}\langle X^{\beta,\alpha_{2}}\rangle\right)}{\omega+E_{\vec{\alpha}}- i\frac{\gamma}{2}}.\label{eq:solforX}
\end{equation}

We have also neglected the contribution of $\langle X \rangle d^{\dagger}\left(\omega\right)$,
as typically done in the literature \cite{PRASigmund}. Substituting this result in (\ref{eq:dadt}), we find

\begin{equation}
d\left(\omega\right)=\frac{i\sum_{l=1,2}\sqrt{\kappa_{l}}b_{\mathrm{in},l}\left(\omega\right)}{\Omega-\omega-i\frac{\kappa}{2}+g^{2}\chi\left(\omega\right)}
\end{equation}

where 

\begin{equation}
\chi\left(\omega\right)=\sum_{\vec{\alpha},\beta}\frac{Z_{\vec{\alpha}} \left(Z_{\alpha_{2},\beta}\langle X^{\alpha_{1},\beta}\rangle-Z_{\beta,\alpha_{1}}\langle X^{\beta,\alpha_{2}}\rangle\right)}{\omega+E_{\vec{\alpha}}- i\frac{\gamma}{2}}.
\label{solforX}
\end{equation}

Using (\ref{eq:inandout}), and taking into account that the input
is inserted through the left port ($l=1$) into the cavity, and the
output is collected through the right one ($l=2$), we can write the
transmission as

\begin{equation}
t_{c}\left(\omega\right)=\frac{\langle b_{\mathrm{out},2}\rangle}{\langle b_{\mathrm{in},1}\rangle}=\frac{i\sqrt{\kappa_{1}\kappa_{2}}}{\Omega-\omega-i\frac{\kappa}{2}+g^{2}\chi\left(\omega\right)}
\end{equation}

which is the usual result for the cavity transmission, with $\chi\left(\omega\right)$
being the electronic susceptibility.\\

\subsubsection*{Using the mean-field Hamiltonian} If we instead consider the MF Hamiltonian for
the cavity, topological system and their interaction, we start from
the following expression

\begin{eqnarray}
H & = & \Omega d^{\dagger}d-\frac{g^{2}\langle Z\rangle^{2}}{\Omega}+\tilde{H}_{S}+g\left(d^{\dagger}+d\right)\tilde{Z} \nonumber\\
&  & +\sum_{l=1,2}\int_{-\infty}^{\infty}\omega b_{l}^{\dagger}\left(\omega\right)b_{l}\left(\omega\right)d\omega \nonumber\\
&  & +i\sum_{l=1,2}\sqrt{\frac{\gamma_{l}}{2\pi}}\int_{-\infty}^{\infty}d\omega\left[b_{l}^{\dagger}\left(\omega\right)\left(d-\frac{g\langle Z\rangle}{\Omega}\right)\right.\nonumber\\
&  & \hspace{1em}\hspace{1em}\hspace{1em}\hspace{1em}\hspace{1em}\hspace{1em}\hspace{1em}\left.-\left(d^{\dagger}-\frac{g\langle Z\rangle}{\Omega}\right)b_{l}\left(\omega\right)\right]
\end{eqnarray}

where $\tilde{Z}=Z-\langle Z\rangle$ and $\tilde{H}_{S}=H_{S}-\frac{2g^{2}\langle Z\rangle}{\Omega}Z$.
Note that the cavity operators have been rotated to $\tilde{d}^{\left(\dagger\right)}=d^{\left(\dagger\right)}-\frac{g\langle Z\rangle}{\Omega}$
in order to diagonalize their MF Hamiltonian. The self-consistent
values of $\langle Z\rangle$ and $\langle d^{\left(\dagger\right)}\rangle$
have been determined ignoring the coupling to the transmission line, as explained in the main text.
Importantly, the cavity operators in $H_{B}$ have also been rotated
accordingly. This time, the EoM for $b_{l}\left(\nu,t\right)$ yields
the following solution

\begin{equation}
\tilde{b}_{l}\left(\nu,t\right) =  b_{l}\left(\nu,t\right)-\frac{g\langle Z\rangle}{\Omega}\sqrt{\frac{\kappa_{l}}{2\pi}}\frac{1-e^{-i\nu\left(t-t_{0}\right)}}{i\nu}
\end{equation}

Compared to (\ref{eq:solforb}), we have an extra term due to the
rotation of the bosonic operators, that depends on the state of the
fermionic system through $\langle Z\rangle$. The EoM for $d\left(t\right)$
has the same form, with redefined input and output fields

\begin{equation}
\tilde{b}_{\mathrm{in},l}\left(t\right) =  b_{\mathrm{in},l}\left(t\right)-\frac{g\sqrt{\kappa_{l}}\langle Z\rangle}{2\pi\Omega}\int_{-\infty}^{\infty}\frac{1-e^{-i\nu\left(t-t_{0}\right)}}{i\nu}d\nu,
\end{equation}

\begin{equation}
\tilde{b}_{\mathrm{out},l}\left(t\right) = b_{\mathrm{out},l}\left(t\right)+\frac{g\sqrt{\kappa_{l}}\langle Z\rangle}{2\pi\Omega}\int_{-\infty}^{\infty}\frac{1-e^{i\nu\left(t_{1}-t\right)}}{i\nu}d\nu,
\end{equation}

also fullfilling that $\tilde{b}_{\mathrm{out},l}\left(t\right)=\tilde{b}_{\mathrm{in},l}\left(t\right)+\sqrt{\kappa_{l}}d\left(t\right)$.
On the other hand, the Hubbard operators change as well due to the
presence of the extra MF contribution in the fermionic Hamiltonian

\begin{eqnarray}
\tilde{H}_{S} & = & H_{S}-\frac{2g^{2}\langle Z\rangle}{\Omega}Z=\sum_{\alpha=1}^{N}\tilde{E}_{\alpha}\tilde{X}^{\alpha,\alpha},\\
\tilde{Z} & = & \sum_{\vec{\alpha}}\tilde{Z}_{\vec{\alpha}}\tilde{X}^{\vec{\alpha}}.
\label{EoMXtilde}
\end{eqnarray}

The EoM for $\tilde{X}$ has the same form of (\ref{eomX}), but substituting the eigenvalues and eigenvectors of the unperturbed fermionic Hamiltonian $H_S$ by the ones of MF Hamiltonian $\tilde{H}_S$. To solve the EoM, we take fluctuations to be small, which is a valid assumption when $g$ is both small and very large, i.e., when the MF Hamiltonian $\tilde{H}_S + \tilde{H}_\Omega$ accurately describes the physics of the system, without considering the fluctuations Hamiltonian $\tilde{H}_\delta$. Under this condition, we can write $	d^{\left(\dagger\right)}\left(t\right)\tilde{X}\left(t\right)\approx\langle \tilde{X} \rangle d^{\left(\dagger\right)}\left(t\right) + \langle d^{\left(\dagger\right)}\rangle \tilde{X}(t)$. Again, we neglect any correlation created by the fluctuations Hamiltonian, which acts as an effective interaction between the two MF Hamiltonians. Note that $\langle d^{(\dagger)} \rangle = 0$ when evaluated using the MF photonic Hamiltonian in the rotated frame, $\tilde{H}_\Omega$. \\
Formally, the decoupling employed to solve (\ref{EoMXtilde}) is the same as the one used in (\ref{eomX}). Then, the solution for $\tilde{X}$ gives

\begin{equation}
\tilde{X}^{\vec{\alpha}}(\omega)\simeq gd\left(\omega\right)\frac{\sum_{\beta}\left(\tilde{Z}_{\alpha_{2},\beta}\langle\tilde{X}^{\alpha_{1},\beta}\rangle-\tilde{Z}_{\beta,\alpha_{1}}\langle\tilde{X}^{\beta,\alpha_{2}}\rangle\right)}{\omega+\tilde{E}_{\vec{\alpha}}-i\frac{\gamma}{2}}.
\label{solforXtilde}
\end{equation}

Note that (\ref{solforX}) is analogous to (\ref{solforXtilde}), but all parameters have been renormalized due to MF. Finally,
the transmission can be written as

\begin{equation}
t_{c}\left(\omega\right) = \frac{i\sqrt{\kappa_{1}\kappa_{2}}}{\Omega-\omega-i\frac{\kappa}{2}+g^{2}\tilde{\chi}\left(\omega\right)}.
\end{equation}

where now $\tilde{\chi}\left(\omega\right)$ is the susceptibility written in terms of the MF Hamiltonian.

\section{Transmission and photonic Green's function}

The starting point is the EoM for the photonic operator $d(t)$ (\ref{eq:dadt}) in Fourier space

\begin{eqnarray}
i\omega d\left(\omega\right) & = & -i\left(\Omega-i\frac{\kappa}{2}\right)d\left(\omega\right)-ig\sum_{\vec{\alpha}}Z_{\vec{\alpha}}\tilde{X}^{\vec{\alpha}}\left(\omega\right) \nonumber \\
& - & \sum_{l=1,2}\sqrt{\kappa_{l}}\tilde{b}_{\mathrm{in},l}\left(\omega\right)
\label{eq:EoMaFourier}
\end{eqnarray}

The losses of the cavity have been included through the phenomenological factor $\kappa$. This equation depends on $\tilde{X}^{\vec{\alpha}}\left(\omega\right)$,
which has its own dynamics as well,

\begin{eqnarray}
i\omega\tilde{X}^{ij} (\omega) & = & -i\left(\tilde{E}_{j} - \tilde{E}_i+i\frac{\gamma}{2}\right)\tilde{X}^{ij}(\omega) \nonumber \\
&  & - ig\left(d^{\dagger}(\omega)  +d(\omega)\right)\sum_{m}\left(Z_{jm}\tilde{X}^{im}(\omega) \right. \nonumber \\
&  & \hspace{60pt} - \left. Z_{mi}\tilde{X}^{mj}(\omega)\right).
\end{eqnarray}

Solving this equation implies writting the EoM for $d^{\left(\dagger\right)}\tilde{X}^{ij}$,
which at the same time is coupled to higher-order operators. Then,
we can write an infinite vector with all the relevant operators involved, $\vec{v}=\left(d,X^{ij}, ...\right)$,
and the system of equations turns out to be

\begin{equation}
\left(\omega- A \right)\vec{v}=\vec{v}_{0}
\end{equation}

where $A$ is the coefficients matrix and $\vec{v}_{0}$ represents
the inhomogeneous term. 

Then, on the other hand, we can write the EoM for the retarded photonic Green
function, defined as $G\left(t\right)=-i\theta\left(t\right)\langle\left[d\left(t\right),d^{\dagger}\right]\rangle \equiv \langle \langle d(t); d^\dagger \rangle \rangle_t $, yielding

\begin{equation}
\omega G\left(\omega\right)=1+\left(\Omega-i\frac{\kappa}{2}\right)\tilde{G}\left(\omega\right)+g\sum_{ij}M_{ij}\left(\omega\right)
\label{eqofMforG}
\end{equation}

where $\tilde{M}_{ij}\left(\omega\right)=\langle\langle X^{ij};a^{\dagger}\rangle\rangle_\omega$. Note that the dissipative factor enters the EoM for $G(\omega)$ through the integration of the external modes and their coupling to the cavity photons, just as in Eq. \ref{eq:dadt}. Again, this EoM is coupled to higher-order Green functions, resulting
in an infinite system of coupled differential equations. In matrix form, we have

\begin{equation}
\left(\omega-A^{\prime}\right)\vec{V}\left(\omega\right)=\vec{V}_{0}
\end{equation}

where $\vec{V}=\left(G,M_{ij},...\right)$ and $\vec{V}_{0}$ is the inhomogeneous
term by comparison. One can see that $A = A^{\prime}$, which indicates  that $G\left(\omega\right) = -i\theta\left(t\right)\langle\left[a\left(t\right),a^{\dagger}\right]\rangle$ is the resolvent of (\ref{eq:EoMaFourier}). 

Finally, we can compare the first component of each system of equations,
namely $a\left(\omega\right)=-i\left(\omega-H\right)^{-1}\sum_{l=1,2}\sqrt{\kappa_{l}}\tilde{b}_{\mathrm{in},l}\left(\omega\right)$
and $G\left(\omega\right)=\left(\omega-H\right)^{-1}$, and see
that

\begin{equation}
a\left(\omega\right)=-iG\left(\omega\right)\sum_{l=1,2}\sqrt{\kappa_{l}}\tilde{b}_{\mathrm{in},l}\left(\omega\right).\label{eq:aasG}
\end{equation}

The last step is to write the transmission as a function of $a\left(\omega\right)$,
knowing that $\tilde{b}_{\mathrm{out,l}}\left(t\right)=\tilde{b}_{\mathrm{in,l}}\left(\omega\right)+\sqrt{\kappa_{l}}a\left(\omega\right)$
and that the only input is through port $1$

\begin{equation}
t_{c}=\frac{\langle \tilde{b}_{\mathrm{out,2}}\rangle}{\langle \tilde{b}_{in,1} \rangle}=\frac{\langle \tilde{b}_{\mathrm{in,2} }\rangle +\sqrt{\kappa_{2}}\langle a \rangle }{\langle \tilde{b}_{in,1}\rangle}=-i\sqrt{\kappa_{2}}\sqrt{\kappa_{1}}G\left(\omega\right).
\end{equation}

It is very enlightening to obtain an analytical expression for the photonic Green function in the case of $g\ll\Omega,\tilde{E}_{\vec{\alpha}}$ and $g\gg\Omega,\tilde{E}_{\vec{\alpha}}$. The equation of motion of $G(\omega)$ (Eq. \ref{eqofMforG}) is coupled to the mixed Green function $\tilde{M}_{ij}$,
which gives

\begin{eqnarray}
\omega \tilde{M}_{ij} & = & (\tilde{E}_j - \tilde{E}_i)\tilde{M}_{ij} + \sum_{l} Z_{jl} \langle \langle (d^\dagger + d) \tilde{X}^{il}; d^\dagger \rangle \rangle_\omega \nonumber \\
& & - \sum_{l} Z_{li} \langle \langle (d^\dagger + d)\tilde{X}^{lj}; d^\dagger \rangle \rangle 
\label{eqofmotionforMij}
\end{eqnarray}

Note the presence of the higher-order Green function $\langle \langle (d^\dagger + d) \tilde{X}; d^\dagger  \rangle \rangle $. Fluctuations are negligible in the small and very-large coupling regimes, which let us employ the following decoupling scheme

\begin{eqnarray}
\langle \langle (d^\dagger + d) \tilde{X}; d^\dagger  \rangle \rangle & \approx & \big( \langle d^\dagger \rangle  + \langle d \rangle \big) \langle \langle \tilde{X};d^\dagger \rangle \rangle_\omega \nonumber \\
& & + \langle \tilde{X} \rangle \langle \langle d ; d^\dagger \rangle \rangle_\omega 
\end{eqnarray}

This approximation is analogous to the decoupling scheme $d^{\left(\dagger\right)}\left(t\right)\tilde{X}\left(t\right)\approx\langle \tilde{X} \rangle d^{\left(\dagger\right)}\left(t\right) + \langle d^{\left(\dagger\right)}\rangle \tilde{X}(t)$ presented in the main text used to solve the EoM for the photonic operator. Again, it implies that we are neglecting first-order correlations between fermionic and photonic operators, and under this assumption the system of equations including Eqs. \ref{eqofMforG} and \ref{eqofmotionforMij} can be closed and solved, obtaining

\begin{equation}
G\left(\omega\right)=-\frac{1}{\Omega-\omega+g^{2}\tilde{\chi}\left(\omega\right)}.
\label{eq:GFsmallg}
\end{equation}

for the photonic Green function. It is straight-forward to see that the previous expression verifies the relation between the transmission and $G(\omega)$ presented in Eq. \ref{eq:Exact-transmission}, when compared with Eq. \ref{eq:Transmission} in the limits of $g\ll\Omega,\tilde{E}_{\vec{\alpha}}$ and $g\gg\Omega,\tilde{E}_{\vec{\alpha}}$.\\

\section{\label{sec:appendixZ}Solution for $\langle Z\rangle$ }

In order to solve the self-consistency equation for the
order parameter $\langle Z\rangle=\sum_{\vec{\alpha}}\tilde{Z}_{\vec{\alpha}} \langle \tilde{X}^{\vec{\alpha}} \rangle$, we iterate until convergence using the eigenstates of the MF Hamiltonian $\tilde{H}_{S}$ in Eq. \ref{fermionicMF}, and also those of the total Hamiltonian $\tilde{H}_\delta + \tilde{H}_S + \tilde{H}_\Omega$ to compare with the exact value. \\

\begin{figure}
	\centering
	\includegraphics{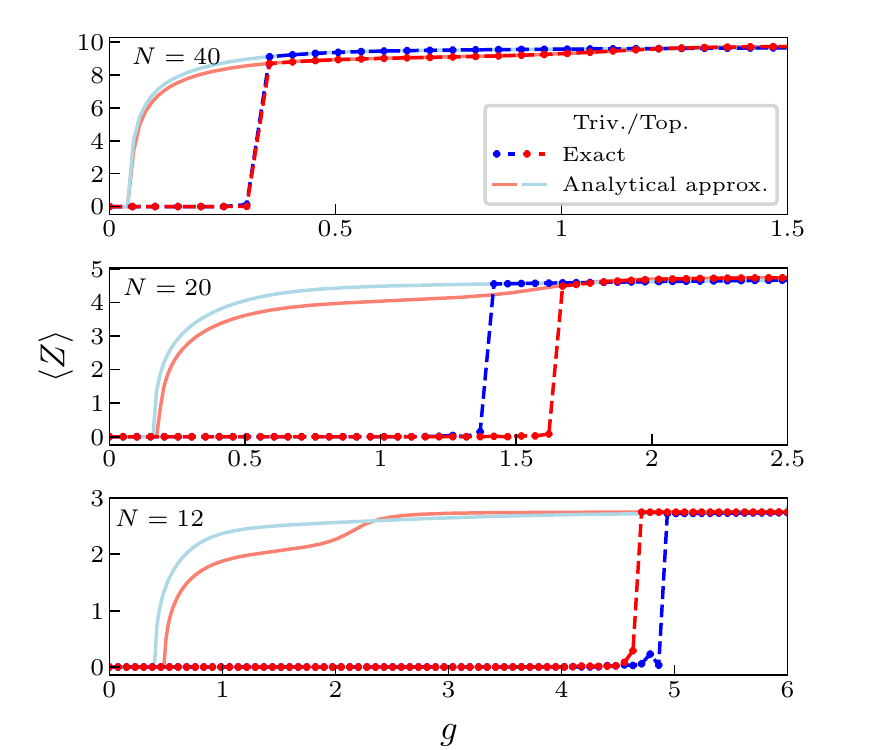}\caption{ $\mathbf{\langle Z \rangle} $ \textbf{for different} $\mathbf{N}$. \label{fig:Zsize}$\langle Z\rangle$ as a function of the coupling constant $g$ for $\Omega=10$, $\delta=\pm0.6$ (trivial/topological phase) and different system sizes: $N=40$ (top), $N=20$ (middle), $N=12$ (bottom). The value of $\langle Z\rangle$ has been calculated self-consistently using the MF Hamiltonian (solid) and exact diagonalization (dashed-dotted line). The chain size modifies the value of the critical point at which the system polarizes, enhancing or reducing the difference between topological phases at the phase transition}
\end{figure}

Fig. \ref{fig:Zsize} shows $\langle Z\rangle$ as a function of $g$
for the MF case and the exact one, and provides us a precise value for the breakdown of the approximation. Different chain lengths $N$ have also been considered. As expected, the MF calculation agrees
with the exact value at small and large $g$. In the former case,
$\langle Z\rangle=0$ indicates that the MF Hamiltonian coincides
with the free Hamiltonian, and that the symmetries of the model are
unaffected by the coupling with the cavity. In the latter, the system
polarizes (i.e., $\langle Z\rangle\neq0$) indicating that the ground
state of the system is modified and that certain symmetries change.
For intermediate values, fluctuations take over and lead
to disagrement between the MF and the exact solution. The MF result
indicates that the change in $\langle Z\rangle$ is continuous, corresponding
to a second-order phase transition. However, the exact calculation
shows a discontinuity at a critical value for $g$, which could indicate that is in fact a first
order transition.\\

Interestingly, we find that beyond the critical $g$ a difference
between the topological and the trivial phase arises, which is also
captured by the MF solution. This is a consequence of the different
coupling between bulk/edge modes and the cavity photons. It also indicates
that certain features of the topological edge states still remain
when the coupling strength is increased beyond the small $g$ regime, but they dissappear again at very large $g$.

$\langle Z \rangle$ also depends on the chain length $N$. This is intuitive due to the position-dependent interaction: the chain size modifies the value of the critical point at which the system polarizes, enhancing or reducing the difference between topological phases at the phase transition, as well as the final value in the limit $g\rightarrow\infty$. Also, as the number of sites $N$ increases, the critical value of $g$ at which the phase transition
happens is reduced (see Fig. \ref{fig:Zsize}). This is expected, since the effective strength of the coupling at each site $g_{i}=gx_{i}$
gets larger as more sites are considered. 

The system size also shapes the differences between topological phases
found inmediatly after the phase transition, that are captured by the
MF calculation, without including fluctuations.




\section{Topological detection in the small coupling regime occupying an edge state}

We have shown that the cavity transmission cannot act as a topological marker if the lowest-energy state is occupied. On the contrary, if the edge state is initially occupied in the topological phase, the transmission peak at $\omega=\Omega$ should remain unaffected by the interaction, as opposed to the behaviour of the trivial phase, in which changes in $t_{c}$ are expected. This assymetry between phases can be maximized if the cavity frequency $\Omega$ is resonant with an electronic transition $E_{\vec{\alpha}}$ (eigenenergy of the unperturbed fermionic Hamiltonian $H_{S}$) and in particular, with the gap of the chain. While the transmission for the non-trivial topological phase does not change compared to the uncoupled cavity transmission (the bulk states are decoupled from the edge states), the presence of a direct resonance in the trivial phase results in a Rabi splitting: the peak of maximum transmission divides into two distinct modes, separated by the Rabi frequency $\Omega_{r}$ \cite{MilburnWallsQO}, which is often detected in experiments and indicates that the regime $g>\{\kappa,\gamma \}$ is achieved \cite{StrongTQDEnsslin,SCofQDResonator,StrongSpinMonicaPetta,StrongSPcouplingSi,StrongCouplingPetta}.  

\begin{figure}[!t]
	\includegraphics{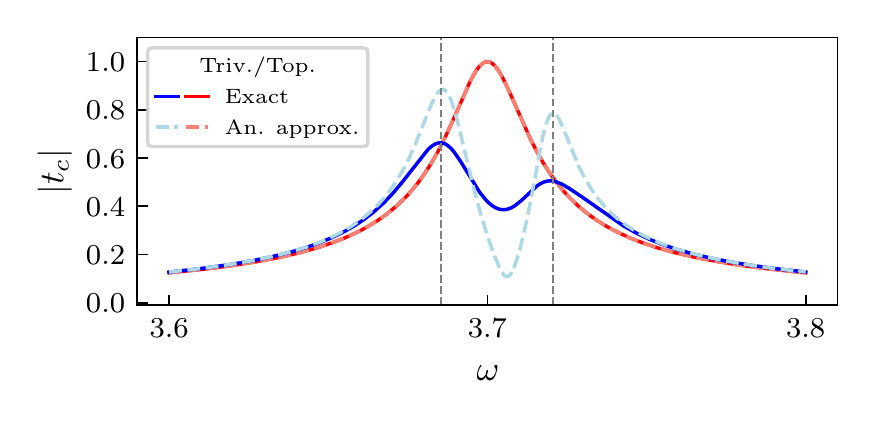}\caption{\label{fig:rabisplitting} $\mathbf{\left|t_{c}\left(\omega\right)\right|}$
		\textbf{as a function of} $\mathbf{\omega}$. Dashed lines indicate
		the analytical approximation (Eq. \ref{eq:Transmission}), while solid
		lines correspond to the exact solution (Eq. \ref{eq:Exact-transmission}).
		We consider the $\left(N/2\right)$-th state occupied, corresponding
		to the edge state in the topological phase and the top of the valence
		band in the trivial phase. The parameters used are:
		$\Omega=3.7$, $g=0.06$, $\delta=\pm0.925$ (trivial/topological
		phase, in blue/red, respectively), $N=20$, $\gamma = \kappa_1 = \kappa_2 = 0.01$.}
\end{figure}

This is shown in Fig. \ref{fig:rabisplitting}. We consider the top state in the valence band is occupied for the trivial phase, while the edge state is occupied for the topological phase. This choice is motivated by the fact that both states are adiabatically connected across the topological phase transition, as $\delta$ is varied from negative to positive values. While the decoupling between bulk and edge states explains the absence of any changes in $t_c$ for the topological phase, a Rabi splitting appears for the trivial phase. The analytical approximation captures the position of each peak in $\omega$ (as indicated by the dashed, grey vertical lines), as well as their relative height, though the exact shape of the peaks is not reproduced. This disagreement can be explained by correlated excitations that strongly modify the system due to resonant conditions. 

In conclusion, for the small-$g$ regime it is important to stress that detection of the topological phase requires the system to be initialized in an edge state. Otherwise the cavity transmission cannot differentiate between the two phases. One disadvantage of this regime of operation is that the measurement needs to be carried out before thermalization happens, but this could be avoided by filling the fermionic system to half-filling (however, in this case many systems require to account for particle interactions as well).

\section{Schrieffer-Wolff transformation}

We begin with the Hamiltonian $H = \tilde{H}_S + \tilde{H}_\Omega + \tilde{H}_\delta$, as defined in the main text, written in the basis of eigenstates of $\tilde{H}_S$:

\begin{equation}
H = \sum_i \tilde{E}_\alpha \tilde{X}^{\alpha,\alpha} + \Omega d^\dagger d - \frac{g^2 \langle Z \rangle }{\Omega}+g(d^\dagger + d) \tilde{Z}
\end{equation}

We propose the following ansatz for the generator $S$ of the Schrieffer-Wolff (S-W) transformation,

\begin{equation}
S = \sum_{\vec{\alpha}} \big( \Gamma^+_{\vec{\alpha}}d^\dagger + \Gamma^{-}_{\vec{\alpha}}d \big)\tilde{X}^{\vec{\alpha}}
\end{equation}

where $\vec{\alpha} = (\alpha_1, \alpha_2)$. Imposing  $\tilde{H}_\delta = - [S, \tilde{H}_S + \tilde{H}_\Omega]$, one finds the following equations for the free parameters:

\begin{equation}
\Gamma^{\pm}_{\vec{\alpha}} = \frac{g \tilde{Z}_{\vec{\alpha}}}{\tilde{E}_{\vec{\alpha}} \pm \Omega}.
\end{equation}

with $\tilde{E}_{\vec{\alpha}} = \tilde{E}_{\alpha_1} - \tilde{E}_{\alpha_2}$. This results in the final form of the transformation:
\begin{equation}
S =  g(d^\dagger + d)\sum_{\vec{\alpha}} \frac{ \tilde{E}_{\vec{\alpha}}  \tilde{Z}_{\vec{\alpha}} }{\tilde{E}_{\vec{\alpha}}^2 - \Omega^2} - g \Omega (d^\dagger - d)  \sum_{ij} \frac{\tilde{Z}_{\vec{\alpha}}}{\tilde{E}_{\vec{\alpha}}^2 - \Omega^2}\tilde{X}^{\vec{\alpha}}
\end{equation}

The correction to the Hamiltonian is proportional to:

\begin{eqnarray}
\left[S,\tilde{H}_\delta \right] & = & g^2 (d^\dagger + d)^2 \sum _{\vec{\alpha}} \frac{\tilde{E}_{\vec{\alpha}} \tilde{Z}_{\vec{\alpha}} }{\tilde{E}_{\vec{\alpha}}^2 - \Omega^2} Y^{-}_{\vec{\alpha}} \nonumber \\
& & -g^2 \Omega (d^{\dagger2} - d^2)\sum_{\vec{\alpha}} \frac{\tilde{Z}_{\vec{\alpha}}}{\tilde{E}_{\vec{\alpha}} ^2 - \Omega^2}Y^{-}_{\vec{\alpha}} \nonumber \\
& & +g^2 \Omega \sum_{\vec{\alpha}} \frac{\tilde{Z}_{\vec{\alpha}}}{\tilde{E}_{\vec{\alpha}}^2 - \Omega^2} Y^{+}_{\vec{\alpha}}
\end{eqnarray}

with

\begin{eqnarray}
Y_{\vec{\alpha}}^{\pm} = \sum_\beta (\tilde{Z}_{\alpha_2 \beta}\tilde{X}^{\alpha_1 \beta} \pm \tilde{Z}_{\beta \alpha_1}X^{\beta \alpha_2})
\end{eqnarray}
Finally, one can write the effective Hamiltonian up to second order, neglecting two-photon processes and constant terms,

\begin{eqnarray}
\bar{H} & \simeq & \tilde{H}_S + \Omega d^\dagger d + \frac{1}{2}[S,\tilde{H}_\delta] \nonumber \\
& = & \sum_{\alpha}\tilde{E}_{\alpha} \tilde{X}^{\alpha,\alpha} \nonumber \\
& & + \frac{g^2}{2}\sum_{\vec{\alpha}} \tilde{Z}_{\vec{\alpha}} \left[ \frac{\sum_\beta \tilde{Z}_{\alpha_2 \beta} \tilde{X}^{\alpha_1 \beta}}{\tilde{E}_{\vec{\alpha}} - \Omega} - \frac{\sum_n \tilde{Z}_{\beta \alpha_1} \tilde{X}^{\beta \alpha_2} }{\tilde{E}_{\vec{\alpha}} + \Omega} \right] \nonumber\\
& & + \bigg[ \Omega + g^2 \sum_{\vec{\alpha}} \tilde{\Omega}_{\vec{\alpha}} \tilde{Y}^{-}_{\vec{\alpha}} \bigg] d^\dagger d.
\end{eqnarray}

where $\tilde{\Omega}_{\vec{\alpha}}$ is defined in the main text.
$\vspace{10pt}$

\section{\label{sec:ent-app}Entanglement Entropy and Energy Spectrum}

As shown in Fig. \ref{fig:entanglement-entropy}(c) in the main text, there is a $\mathrm{log}\left(2\right)$-plateau
in the topological phase for small-$g$ when the $N/2$-th state is
occupied, corresponding to one of the topological edge states. Its
drop at $g\sim0.84$ corresponds with its anti-crossing with a state
belonging to the bulk bands and indicates the destruction of maximal
entanglement for the $N/2$-th state.

\begin{figure*}[h]
	\includegraphics{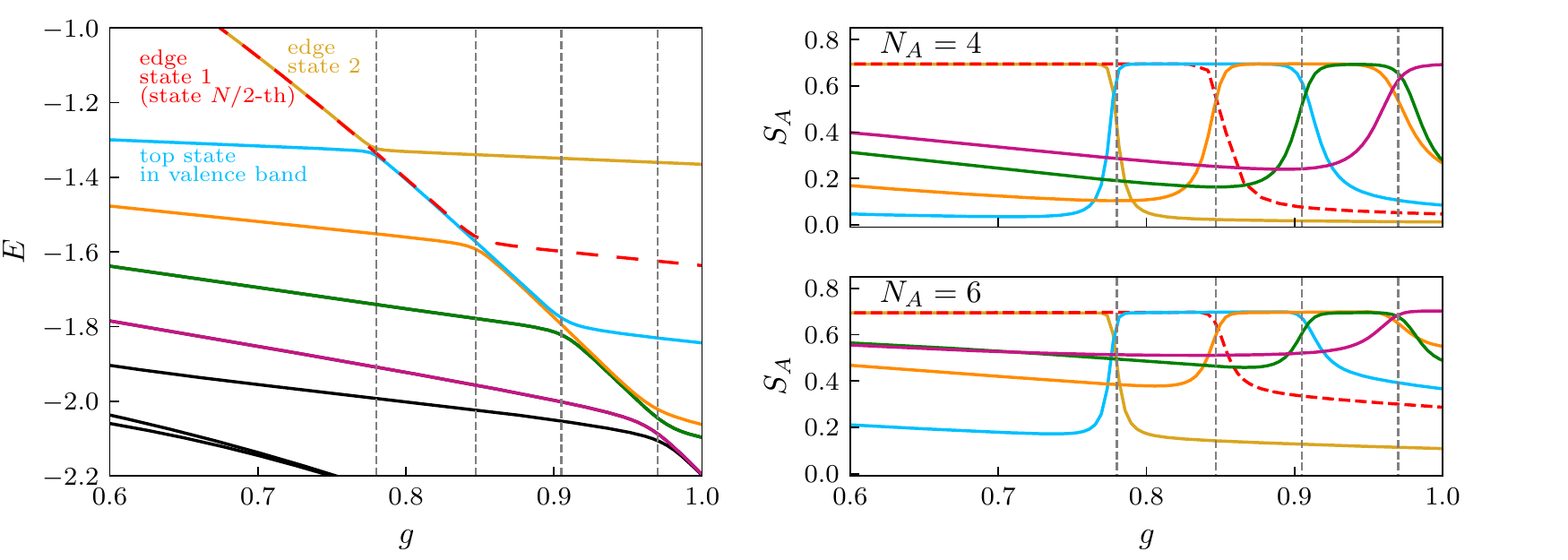}
	\caption{\label{fig:to-entalg} \textbf{Entanglement entropy and the energy spectrum}. Left plot: zoom into the energy spectrum, as
		a function of $g$, corresponding to the anti-crossing (dashed, grey
		vertical lines) of energy states triggered by the entrance of the
		edge states into the bulk band. The parameters chosen are: $\Omega=10$
		$\delta=-0.6$, $N=20$. Each state is depicted in a different color:
		light brown for the $\left(\frac{N}{2}+1\right)$ -th state, red for the
		$N/2$-th (corresponding to the two edge states), blue for the $\left(\frac{N}{2}-1\right)$-th
		(top state in the valence band), orange for the $\left(\frac{N}{2}-2\right)$-th,
		green for the $\left(\frac{N}{2}-3\right)$-th and violet for the
		$\left(\frac{N}{2}-4\right)$-th (the following states appear in black).
		The first anti-crossing at $g\sim0.78$ happens between the edge state
		(light brown) and the top state in the valence band (blue). The second
		anti-crossing at $g\sim0.9$ happens between the edge state (red)
		and $\left(\frac{N}{2}-2\right)$-th state in the bulk band (orange).
		For higher values of the coupling constant (only shown up to $g=1$),
		sucessive anti-crossings between adjacent states appear. Right plot:
		entanglement entropy for each state (same color code as in the left
		plot), for different partitions $N=4$ (upper plot) and $N=6$ (lower
		plot). For small-$g$, the $\mathrm{log}\left(2\right)$ plateau corresponds
		to the $N/2$-th state, while the rest of them are not maximally entangled
		(their $S_{A}$ depends on the partition used). Each time a state
		from the bulk band anti-crosses with an edge state, it turns into
		an edge state itself, so that the $\mathrm{log}\left(2\right)$ plateau
		(originally caused by the non-trivial topology of the fermionic system)
		migrates from one state to the other.}
\end{figure*}

However, the topological contribution to the entanglement entropy
is not lost at this point, but migrates from one state to the other
as they further anti-cross in the energy spectrum.
Figure \ref{fig:to-entalg} shows a zoom of the energy spectrum where
the edge states penetrate into the bulk bands, together with the entanglement
entropy $S_{A}$ associated to the occupation of each of them. For
small-$g$, the $\mathrm{log}\left(2\right)$ plateau corresponds
to the $N/2$-th (red) and $\left(N/2+1\right)$-th (light brown) states
(edge states), while the rest of them are not maximally entangled
(their $S_{A}$ depends on the partition used). The first anti-crossing
encountered in the spectrum (at $g\sim0.78$) between the light brown and
blue (top state in the valence band) states, corresponds to the appearance
of a $\mathrm{log}\left(2\right)$-plateau for the later, while the entanglement between the ending sites is lost for the edge state.

\begin{figure*}
	\includegraphics{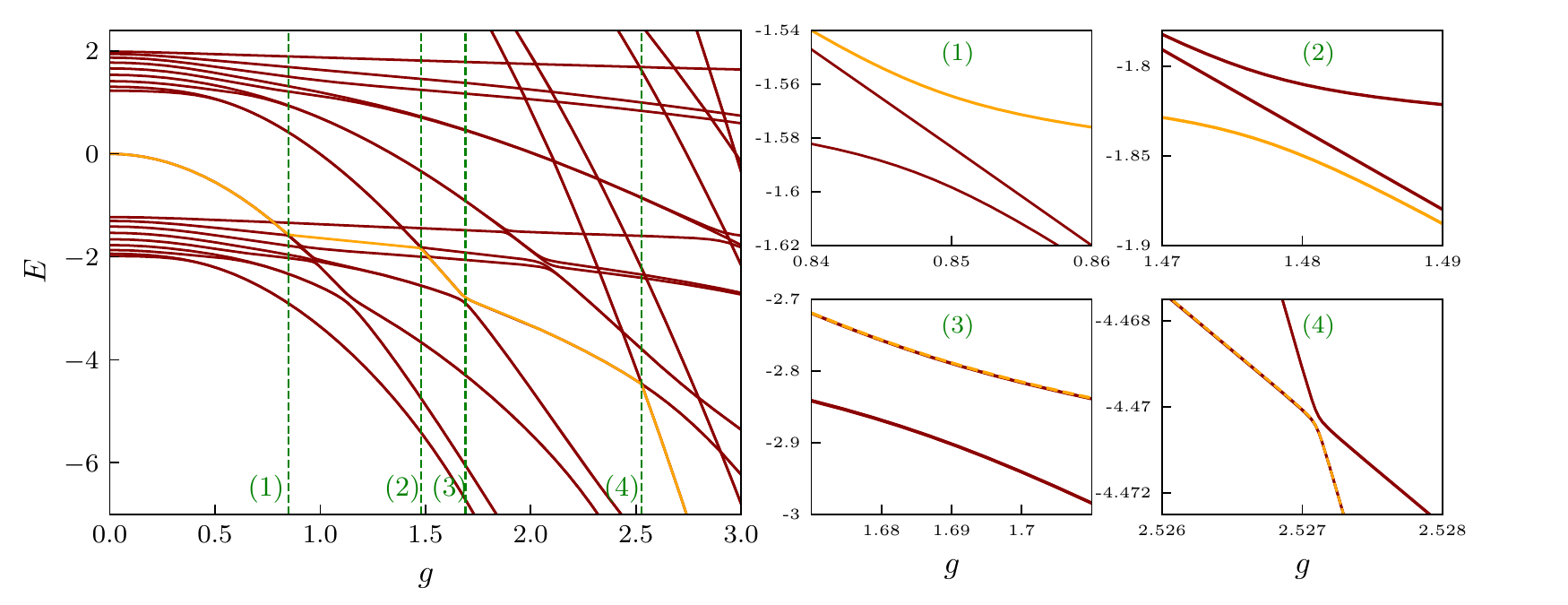}
	
	\caption{\label{fig:anticrossings} \textbf{Anti-crossings for the} $\mathbf{N/2^\mathrm{th}}$ \textbf{state in the topological phase.} Left plot: energy spectrum for the zero-photon	band as a function of $g$. The $N/2$-th state (for which the entanglement entropy is calculated in Fig. 6(c)) is represented in orange. The green, dashed lines represent the first four anti-crossings for this state, which correspond to the first four abrupt changes in $S_{A}$ in Fig. 6(c). Right plots: zoom into the first four anti-crossings of the $N/2$-th state.} 
	
\end{figure*}

On the other hand, as $g$ is increased, the $N/2$-th state anti-crosses
with other states as well. Figure \ref{fig:anticrossings} shows that
each of these anti-crossings correspond to the abrupt changes in $S_{A}$
obtained for the $N/2$-th state (Fig. \ref{fig:entanglement-entropy}(c)),
by zooming into the first four of them.

\end{document}